\documentclass[superscriptaddress,onecolumn,pre,nofootinbib]{revtex4}

\usepackage{amsmath}
\usepackage{graphicx,color}

\newcommand{\be}{\begin{equation}}
\newcommand{\ee}{\end{equation}}
\newcommand{\bea}{\begin{eqnarray}}
\newcommand{\eea}{\end{eqnarray}}

\begin{document}
\sloppy

%-title page-%

\title{Initial value problem for the linearized mean field Kramers equation\\
with long-range interactions}

\author{Pierre-Henri Chavanis}
%\email{chavanis@irsamc.ups-tlse.fr}
\affiliation{Laboratoire de Physique Th\'eorique (IRSAMC), CNRS and UPS, Universit\'e de Toulouse, F-31062 Toulouse, France}

\begin{abstract}

We solve the initial value problem for the linearized mean field
Kramers equation describing Brownian particles with long-range
interactions in the $N\rightarrow +\infty$ limit. We show that the
dielectric function can be expressed in terms of incomplete Gamma
functions.  The dielectric functions associated with the linearized
Vlasov equation and with the linearized mean field Smoluchowski
equation are recovered as special cases corresponding to the no
friction limit or to the strong friction limit respectively. Although  
the stability of the
Maxwell-Boltzmann distribution is independent on the friction
parameter, the evolution of the perturbation depends on it in a
non-trivial manner. For illustration, we apply our results to self-gravitating
systems, plasmas, and to the attractive and repulsive BMF models.

\end{abstract}

\maketitle

\section{Introduction}
\label{sec_model}

The statistical mechanics of systems with long-range interactions is
currently a topic of active research
\cite{houches,assisebook,oxford,cdr}. In most papers devoted to this
subject, one assumes that the system is isolated. This corresponds to
the {\it microcanonical ensemble} in which the energy is
conserved. This is the correct description of plasmas, stellar
systems, and two-dimensional vortices
\cite{balescubook,bt,paddy,newton,chavhouches,ijmpb,bv}. This is
also the correct description of the Hamiltonian mean field (HMF) model
\cite{ar} which is a toy model of systems with long-range interactions
consisting in $N$ particles of unit mass moving on a circle and
interacting via a cosine potential. In the collisionless regime, valid
for $N\rightarrow +\infty$ in a proper thermodynamic limit, Hamiltonian 
systems with long-range interactions are described by the Vlasov equation. The dynamical stability
of a spatially homogeneous steady state of the Vlasov equation has
been studied by Landau
\cite{landau} in a seminal paper by solving an initial value problem
(previous treatments by Vlasov \cite{vlasov1,vlasov2} and others were
not rigorous and led to mathematical difficulties). 
For the Coulombian
potential, Landau showed that the density perturbation exhibits a
phenomenon of collisionless damping.\footnote{Recently, Mouhot
and Villani \cite{villani}
have obtained an important theorem concerning the nonlinear Landau
damping.}
For the gravitational potential,
the density perturbation either decays or grows depending on whether
the wavelength of the perturbation is smaller or larger than the Jeans
length \cite{lb62,nyquistgrav}. 

In many situations of physical interest, the system is not isolated
from the surrounding and it is important to take into account its
interaction with the external medium. This interaction usually results
in some effects of forcing and dissipation. In the simplest situation,
the one that we shall consider here, the forcing and the dissipation
satisfy a detailed balance condition so that formally the system can
be thought to be in contact with a thermal bath fixing its temperature
$T$. In that case, the proper statistical ensemble is the {\it
canonical ensemble}. We stress that the thermostat is played by a
system of another nature (physically different from
the system under consideration) which usually has short-range
interactions\footnote{Indeed, it is not possible to define the notion
of thermostat for a purely long-range system (i.e. to divide the
system into a subsystem $+$ a reservoir) since the energy is
non-additive \cite{cdr}.}. We shall consider a system of Brownian
particles in interaction for which the deterministic Hamiltonian equations are replaced by stochastic
Langevin equations \cite{hb1,hb2,hb5,longshort}.  In addition to the long-range interaction, the particles experience a friction force and a stochastic force (noise).
If we assume a detailed balance condition, the diffusion coefficient
$D$ and the friction coefficient $\xi$ satisfy the Einstein relation
$D=\xi k_B T/m$ where $T$ is the temperature
of the bath. The self-gravitating Brownian gas has been
studied in a series of papers by Chavanis and Sire (see, e.g.,
\cite{virial1} and references therein) in the strong friction limit
$\xi\rightarrow +\infty$ in which the motion of the particles is
overdamped. Some interesting analogies with the chemotaxis of
bacterial populations, the so-called Keller-Segel model \cite{ks},
have been developed in these papers. The gravitational collapse of the
self-gravitating Brownian gas also presents striking analogies with
the Bose-Einstein condensation (in particular, a Dirac peak is formed
in the post-collapse regime) \cite{sopikbec}. Another example of Brownian systems with long-range
interactions is the Brownian mean field (BMF) model
\cite{cvb,bco,cbo,bmf} which can be viewed as the canonical
counterpart of the HMF model. In the collisionless regime, valid for
$N\rightarrow +\infty$ in a proper thermodynamic limit, Brownian  systems
with long-range interactions 
are described by the mean field Kramers equation. In this paper, we
solve the initial value problem for the linearized mean field
Kramers equation around the spatially homogeneous Maxwell-Boltzmann
distribution.  We obtain the exact solution of this problem and determine the
corresponding dielectric function. We show that it can be expressed in
terms of incomplete Gamma functions.  The zeros of the dielectric
function determine the complex pulsations of the density
perturbations depending on the temperature $T$, the wavenumber $k$, and the
friction coefficient $\xi$. We show that the stability of the spatially
homogeneous Maxwell-Boltzmann distribution is independent on the
friction coefficient. By contrast, the complex pulsations that determine the evolution of the perturbation depend on it
in a non-trivial manner. For $\xi\rightarrow 0$ (no friction limit) we
recover the results of the Vlasov equation and for $\xi\rightarrow +\infty$
(strong friction limit) we recover the results of the Smoluchowski
equation. These results are illustrated for self-gravitating systems, plasmas,
and for the attractive and repulsive BMF models.

\section{Brownian particles in interaction: inertial model}
\label{sec_inertial}

\subsection{The Langevin equations}
\label{sec_lang}

We consider a system of $N$ Brownian particles in interaction. The
dynamics of these particles is governed by the coupled stochastic
Langevin equations
\begin{eqnarray}
{d{\bf r}_{i}\over dt}&=&{\bf v}_{i}, \qquad\qquad\qquad\qquad\qquad\qquad\qquad\quad\nonumber\\
{d{\bf v}_{i}\over dt}&=&-\frac{1}{m}\nabla_i U({\bf r}_{1},...,{\bf r}_{N})-\xi {\bf v}_{i}+\sqrt{2D}{\bf R}_{i}(t).
\label{bkram1}
\end{eqnarray}
The particles interact through
the potential $U({\bf r}_{1},...,{\bf
r}_{N})=\sum_{i<j}m^2 u(|{\bf r}_{i}-{\bf r}_{j}|)$. The
Hamiltonian is $H=\sum_{i=1}^{N}m{v_{i}^{2}/2}+U({\bf
r}_{1},...,{\bf r}_{N})$.  ${\bf
R}_i(t)$ is a Gaussian white noise satisfying $\langle {\bf
R}_i(t)\rangle={\bf 0}$ and $\langle
R_i^{\alpha}(t)R_j^{\beta}(t')\rangle=\delta_{ij}\delta_{\alpha\beta}\delta(t-t')$ where $i=1,...,N$ label the particles and $\alpha=1,...,d$ the coordinates of space. $D$
and $\xi$ are respectively the diffusion and friction
coefficients. The former measures the strength of the noise, whereas
the latter quantifies the dissipation to the external environment. We
assume that these two effects have the same physical origin, like when
the system interacts with a heat bath. In particular, we suppose that
the temperature $T$ of the bath satisfies the Einstein relation
\begin{equation}
D=\frac{\xi k_B T}{m}.
\label{einstein1}
\end{equation}
The temperature measures the strength of the stochastic force for
a given friction coefficient. For $\xi=D=0$, we recover the
Hamiltonian equations of particles in interaction which conserve the
energy $E=H$.

\subsection{The $N$-body Kramers equation}
\label{sec_bkram}

The evolution of the $N$-body distribution function is governed by the Fokker-Planck  equation \cite{hb2}:
\begin{eqnarray}
{\partial P_{N}\over\partial t}+\sum_{i=1}^{N}\biggl ({\bf v}_i\cdot {\partial P_{N}\over\partial{\bf r}_{i}}+{\bf F}_{i}\cdot {\partial P_{N}\over\partial {\bf v}_i}\biggr )=
\sum_{i=1}^{N}{\partial\over\partial {\bf v}_i}\cdot \biggl (D{\partial P_{N}\over\partial {\bf v}_i}+ \xi P_{N}{\bf v}_i\biggr ),
\label{bkram5}
\end{eqnarray}
where ${\bf F}_{i}=-\frac{1}{m}\nabla_iU$ is the force per unit mass acting on particle $i$. This is the so-called $N$-body Kramers equation. In the absence of forcing and dissipation ($\xi=D=0$), it reduces to the Liouville equation. The  $N$-body Kramers equation satisfies an $H$-theorem for the free energy
\begin{equation}
\label{bkram5b}  F[P_N]=E[P_N]-TS[P_N],
\end{equation}
where $E[P_N]=\int P_N H \, d{\bf r}_1 d{\bf v}_1...d{\bf r}_N d{\bf v}_N$ is the energy and  $S[P_N]=-k_B\int P_N \ln P_N \, d{\bf r}_1 d{\bf v}_1...d{\bf r}_N d{\bf v}_N$ is the entropy. A simple calculation gives
\begin{eqnarray}
\dot F=-\sum_{i=1}^N\int\frac{\xi m}{P_N}\left (\frac{k_B T}{m}\frac{\partial P_N}{\partial {\bf v}_i}+P_N {\bf v}_i\right )^2\, d{\bf r}_1 d{\bf v}_1...d{\bf r}_N d{\bf v}_N.
\label{bkram6}
\end{eqnarray}
Therefore, $\dot F\le 0$ and $\dot F=0$ if, and only, if $P_N$ is the canonical distribution defined by Eq.  (\ref{mcd3}) below. Because of the $H$-theorem, the system converges towards the canonical distribution for $t\rightarrow +\infty$.

\subsection{The canonical distribution}
\label{sec_mcd}

When the system is in contact with a thermal bath, the relevant
statistical ensemble is the canonical ensemble. The statistical
equilibrium state is described by the canonical distribution
\begin{eqnarray}
P_{N}({\bf r}_{1},{\bf v}_{1},...,{\bf r}_{N},{\bf v}_{N})={1\over Z(\beta)}e^{-\beta H({\bf r}_{1},{\bf v}_{1},...,{\bf r}_{N},{\bf v}_{N})},
\label{mcd3}
\end{eqnarray}
where $\beta=1/(k_B T)$ is the inverse temperature and $Z(\beta)=\int e^{-\beta H}\, \prod_{i}d{\bf r}_{i}d{\bf v}_i$ is the partition function determined by the normalization condition
$\int P_N  \, \prod_{i}d{\bf r}_{i}d{\bf v}_i=1$. The
canonical distribution (\ref{mcd3}) is the steady state of the
$N$-body Kramers equation (\ref{bkram5}) provided that the Einstein relation (\ref{einstein1}) is satisfied.

We define the free energy by $F(T)=-k_B T\ln Z(T)$. We also introduce the Massieu function $J(\beta)=-\beta F(\beta)=\ln Z(\beta)$. In the canonical ensemble, the average energy $E=\langle H\rangle$ is given by $E=\partial (\beta F)/\partial\beta=-\partial J/\partial\beta$. The fluctuations of energy are given by $\langle H^2\rangle-\langle H\rangle^2=k_B T^2 C$ where $C=dE/dT$ is the specific heat. This relation implies that the specific heat is always positive in the canonical ensemble.

We note that the canonical distribution (\ref{mcd3}) is the minimum of $F[P_N]$ respecting the normalization condition. At equilibrium, we get  $F[P_N]=-k_B T\ln Z(T)=F(T)$.

\subsection{The mean field Kramers equation}

In a proper thermodynamic limit $N\rightarrow +\infty$, we can neglect the correlations between the particles \cite{cdr}. Therefore, the mean field approximation is exact and the $N$-body distribution function can be factorized in a product of $N$ one-body distribution functions
\begin{equation}
\label{mfa1}
P_N({\bf r}_1,{\bf v}_1,...,{\bf r}_N,{\bf v}_N,t)=\prod_{i=1}^{N}P_1({\bf r}_i,{\bf v}_i,t).
\end{equation}
Substituting this factorization in Eq. (\ref{bkram5}) and integrating over $N-1$ variables we find that the evolution of the distribution function $f({\bf r},{\bf v},t)=Nm P_1({\bf r},{\bf v},t)$ is governed by the mean field  Kramers equation \cite{hb2}:
\begin{equation}
\label{browbbgky6}
{\partial f\over\partial t}+{\bf v}\cdot {\partial f\over\partial{\bf r}}-\nabla\Phi\cdot {\partial f\over\partial {\bf v}}={\partial \over\partial {\bf v}}\cdot \biggl (D{\partial f\over\partial {\bf v}}+\xi f{\bf v}\biggr ),
\end{equation}
where
\begin{eqnarray}
\Phi({\bf r},t)=\int u({\bf r}-{\bf r}')\rho({\bf r}',t)\, d{\bf r},
\label{smf10}
\end{eqnarray}
is the mean potential and $\rho({\bf r},t)=\int f({\bf r},{\bf v},t)\, d{\bf v}$ is the density.  For $\xi=D=0$, Eq. (\ref{browbbgky6}) reduces to the Vlasov equation which describes the collisionless evolution of a Hamiltonian system with long-range interactions. Using the Einstein relation (\ref{einstein1}), the mean field Kramers equation (\ref{browbbgky6}) may  be rewritten as
\begin{equation}
\frac{\partial f}{\partial t}
+{\bf v}\cdot \frac{\partial f}{\partial{\bf r}}
-\nabla\Phi\cdot
\frac{\partial f}{\partial {\bf v}}
=
\frac{\partial}{\partial {\bf v}}\cdot \left\lbrack \xi\left (\frac{k_B T}{m}\frac{\partial f}{\partial {\bf v}}+f {\bf v}\right )\right\rbrack.
\label{browbbgky7}
\end{equation}
The mean field Kramers equation satisfies an $H$-theorem for the free energy
\begin{eqnarray}
\label{ffg}
F[f]=E[f]-T S[f]=\frac{1}{2}\int f v^2\, d{\bf r} d{\bf v}+\frac{1}{2}\int\rho\Phi\, d{\bf r}+k_B T\int \frac{f}{m}\ln \left (\frac{f}{Nm}\right )\, d{\bf r} d{\bf v}.
\end{eqnarray}
Its expression can be obtained from Eq. (\ref{bkram5b}) by using the mean field approximation (\ref{mfa1}). In terms of the free energy, the mean field Kramers equation may be written as a gradient flow
\begin{equation}
\frac{\partial f}{\partial t}
+{\bf v}\cdot \frac{\partial f}{\partial{\bf r}}
-\nabla\Phi\cdot
\frac{\partial f}{\partial {\bf v}}
=\frac{\partial}{\partial {\bf v}}\cdot \left \lbrack \xi f\frac{\partial}{\partial {\bf v}}\left (\frac{\delta F}{\delta f}\right )\right\rbrack.
\label{browbbgky7b}
\end{equation}
A simple calculation gives
\begin{equation}
{\dot F}=-\int \xi f\left\lbrack\frac{\partial}{\partial {\bf v}}\left (\frac{\delta F}{\partial f}\right )\right\rbrack^2\, d{\bf r} d{\bf v}=-\int\frac{\xi}{f}\left (\frac{k_BT}{m}\frac{\partial f}{\partial {\bf v}}+f{\bf v}\right )^2\, d{{\bf r}}d{\bf v}.
\label{browbbgky7bb}
\end{equation}
Therefore, $\dot F\le 0$ and $\dot F=0$ if, and only if, $f$ is the mean field
Maxwell-Boltzmann distribution
\begin{equation}
\label{mba1}
f({\bf r},{\bf v})=A\, e^{-\beta m \lbrack \frac{v^2}{2}+\Phi({\bf r})\rbrack},
\end{equation}
with the temperature of the bath $T$. Because of the $H$-theorem, the system converges, for $t\rightarrow +\infty$, towards a mean-field Maxwell-Boltzmann distribution that is a (local) minimum of free energy at fixed mass. If several minima exist at the same temperature, the selection depends on a notion of basin of attraction.  The relaxation time is $t_{B}\sim 1/\xi$, independent of $N$.

\section{Brownian particles in interaction: overdamped model}
\label{sec_obmf}

\subsection{The Langevin equations}
\label{sec_overlang}

In the strong friction limit $\xi\rightarrow +\infty$, the inertia of
the particles can be neglected. This corresponds to the overdamped
model. The stochastic Langevin equations (\ref{bkram1}) reduce to
\begin{eqnarray}
{d{\bf r}_{i}\over dt}=-\mu \nabla_iU({\bf r}_{1},...,{\bf r}_{N})+\sqrt{2D_{*}}{\bf R}_{i}(t),
\label{smf1}
\end{eqnarray}
where $\mu=1/(\xi m)$ is the mobility and $D_{*}=D/\xi^{2}$ is the
diffusion coefficient in physical space. The Einstein relation
(\ref{einstein1}) may be rewritten as
\begin{eqnarray}
D_{*}=\frac{k_B T}{\xi m}=\mu k_B T.
\label{einstein2}
\end{eqnarray}
The temperature measures the strength of the stochastic force (for a given mobility).

\subsection{The $N$-body Smoluchowski equation}
\label{sec_smf}

The evolution of the
$N$-body distribution function $P_N({\bf r}_1,...,{\bf r}_N,t)$ is governed by the $N$-body
Fokker-Planck equation \cite{hb2}:
\begin{eqnarray}
{\partial P_{N}\over\partial t}=\sum_{i=1}^{N}{\partial\over\partial {\bf r}_{i}}\cdot \biggl\lbrack D_{*}{\partial P_{N}\over\partial{\bf r}_{i}}+\mu P_{N}{\partial\over\partial {\bf r}_{i}}U({\bf r}_{1},...,{\bf r}_{N})\biggr\rbrack.\qquad
\label{smf3}
\end{eqnarray}
This is the so-called $N$-body Smoluchowski equation. It can be derived directly from the stochastic equations (\ref{smf1}). Alternatively, it can be obtained from the $N$-body Kramers equation (\ref{bkram5}) in the strong friction limit $\xi\rightarrow +\infty$ \cite{risken}. In that limit, using the Einstein relation (\ref{einstein1}), we find that
\begin{eqnarray}
\label{smf4}
P_N({\bf r}_1,{\bf v}_1,...,{\bf r}_N,{\bf v}_N,t)=\left (\frac{\beta m}{2\pi}\right )^{dN/2} P_N({\bf r}_1,...,{\bf r}_N,t) e^{-\beta m \sum_{i=1}^{N}\frac{{v}_i^{2}}{2}}+O(\xi^{-1}),\qquad\qquad
\end{eqnarray}
where the evolution of $P_N({\bf r}_1,...,{\bf r}_N,t)$ is governed by Eq. (\ref{smf3}).
The $N$-body Smoluchowski equation satisfies an H-theorem for the free energy
\begin{eqnarray}
\label{smf5}
F[P_N]=\int P_N U\, d{\bf r}_1...d{\bf r}_N+k_BT\int P_N\ln P_N\, d{\bf r}_1...d{\bf r}_N
-\frac{d}{2}Nk_B T\ln\left (\frac{2\pi k_B T}{m}\right ).\qquad
\end{eqnarray}
The expression (\ref{smf5}) can be obtained from the free energy (\ref{bkram5b})  by using Eq. (\ref{smf4}). A simple calculation gives
\begin{eqnarray}
\label{smf6}
\dot F=-\sum_{i=1}^{N}\int \frac{m}{\xi P_N}\left (\frac{k_B T}{m}{\partial P_{N}\over\partial{\bf r}_{i}}+\frac{1}{m} P_{N}{\partial U\over\partial {\bf r}_{i}}\right )^2\, d{\bf r}_1...d{\bf r}_N.
\end{eqnarray}
Therefore, $\dot F\le 0$ and $\dot F=0$ if, and only, if $P_N$ is the canonical distribution in physical space defined by Eq. (\ref{smf7}) below. Because of the $H$-theorem, the system converges towards the canonical distribution (\ref{smf7}) for $t\rightarrow +\infty$.

\subsection{The canonical distribution}
\label{sec_mcdb}

The statistical equilibrium state in configuration space is described by the canonical distribution
\begin{eqnarray}
P_{N}({\bf r}_{1},...,{\bf r}_{N})={1\over Z_{conf}(\beta)}e^{-\beta U({\bf r}_{1},...,{\bf r}_{N})},
\label{smf7}
\end{eqnarray}
where $Z_{conf}(\beta)=\int e^{-\beta U}\, \prod_{i}d{\bf r}_{i}$
is the configurational partition function determined by the normalization condition $\int P_N \, d{\bf r}_1....d{\bf r}_N=1$. The canonical distribution (\ref{smf7})  is the steady state of the $N$-body Smoluchowski equation (\ref{smf3}) provided that the Einstein relation (\ref{einstein2}) is satisfied. It can also be obtained from Eq. (\ref{mcd3}) by integrating over the velocity. We then find that $Z(\beta)=Z_{conf}(\beta)(2\pi/\beta m)^{dN/2}$.

We note that the   canonical distribution (\ref{smf7}) is the minimum of $F[P_N]$ respecting the normalization condition. At equilibrium, we get  $F[P_N]=-k_B T\ln Z_{conf}(T)-\frac{d}{2}N k_B T\ln(2\pi k_B T/m)=-k_B T\ln Z(T)=F(T)$.

\subsection{The mean field Smoluchowski equation}
\label{sec_mfse}

In a proper thermodynamic limit $N\rightarrow +\infty$, we can neglect the
correlations between the particles \cite{cdr}. Therefore, the mean field
approximation is exact and the $N$-body distribution function can be
factorized in a product of $N$ one-body distribution functions
\begin{eqnarray}
P_N({\bf r}_1,...,{\bf r}_N,t)=P_1({\bf r}_1,t)...P_1({\bf r}_N,t).\label{smf9b}
\end{eqnarray}
Substituting this factorization in Eq. (\ref{smf3}) and integrating over $N-1$ variables we find that the evolution of the density $\rho({\bf r},t)=N m P_1({\bf r},t)$ is governed by the mean field Smoluchowski equation \cite{hb2}:
\begin{eqnarray}
{\partial\rho\over\partial
t}=\nabla \cdot \left\lbrack {1\over\xi} \biggl
(\frac{k_B T}{m}\nabla\rho+\rho\nabla\Phi \biggr
)\right\rbrack, \label{smf10g}
\end{eqnarray}
where $\Phi({\bf r},t)$ is given by Eq. (\ref{smf10}). The mean field
Smoluchowski equation (\ref{smf10g}) can also be obtained from the
mean field Kramers equation (\ref{browbbgky7}) in the strong friction limit $\xi\rightarrow +\infty$ \cite{risken}.
In that limit, the distribution function is
close to the Maxwellian
\begin{equation}
\label{smf11}
f({\bf r},{\bf v},t)=\left (\frac{\beta m}{2\pi}\right )^{d/2}\rho({\bf r},t)
e^{-\beta m{v^{2}\over 2}}+O(\xi^{-1}),
\end{equation}
with the temperature of the bath, and the evolution of the density is governed by Eq. (\ref{smf10g}).
The mean field Smoluchowski equation (\ref{smf10g}) may be written in the form of an integro-differential equation as
\begin{eqnarray}
\xi{\partial\rho\over\partial t}=\frac{k_B T}{m}\Delta \rho+\nabla\cdot\left\lbrack \rho\nabla\int u({\bf r}-{\bf r}')\rho({\bf r}',t)\, d{\bf r}'\right\rbrack.
\label{dyn1}
\end{eqnarray}
It satisfies an $H$-theorem for the free energy
\begin{eqnarray}
\label{tsce1zero}
F[\rho]={1\over 2}\int\rho\Phi \, d{\bf r}+k_B T\int \frac{\rho}{m}\ln\left (\frac{\rho}{Nm}\right )\, d{\bf r}-\frac{d}{2}N k_B T\ln \left (\frac{2\pi k_B T}{m}\right ).
\end{eqnarray}
The expression (\ref{tsce1zero}) can be obtained from Eq. (\ref{smf5}) by using the mean field approximation (\ref{smf9b}). It can also be obtained from Eq. (\ref{ffg}) by using Eq. (\ref{smf11}). In terms of the free energy, the mean field Smoluchowski equation may be written as a gradient flow
\begin{eqnarray}
{\partial\rho\over\partial t}=\nabla\cdot \left\lbrack\frac{\rho}{\xi}\nabla \left (\frac{\delta F}{\delta\rho}\right )\right\rbrack.
\label{gflow}
\end{eqnarray}
A simple calculation gives
\begin{equation}
\label{smf13}
\dot F=-\int\frac{\rho}{\xi}\left\lbrack\nabla\left (\frac{\delta F}{\partial\rho}\right )\right\rbrack^2\, d{\bf r}=-\int \frac{1}{\xi \rho}\left (\frac{k_BT}{m}\nabla\rho+ \rho \nabla\Phi\right )^2\, d{\bf r}.
\end{equation}
Therefore, $\dot F\le 0$ and $\dot F=0$ if, and only if, $\rho$ is the
mean field Boltzmann distribution
\begin{eqnarray}
\label{cp4}
\rho({\bf r})=A'\, e^{-\beta m \Phi({\bf r})},
\end{eqnarray}
with the temperature of the bath $T$. This distribution can also be obtained from the mean field Maxwell-Boltzmann distribution (\ref{mba1}) by integrating over the velocity. Because of the $H$-theorem, the system converges, for $t\rightarrow +\infty$, towards a mean-field
Boltzmann distribution that is a (local) minimum of free energy
at fixed mass.\footnote{The steady states of
the mean field Smoluchowski equation are the critical points (minima,
maxima, saddle points) of the free energy $F[\rho]$ at fixed mass.
It can be shown \cite{nfp} that a critical
point of free energy is dynamically stable with respect to the mean
field Smoluchowski equation if, and only if, it is a (local)
minimum. Maxima are unstable for all perturbations so they cannot be
reached by the system. Saddle points are unstable only for certain
perturbations so they can be reached if the system does not
spontaneously generate these dangerous perturbations. The same
comments apply to the mean field Kramers equation
(\ref{browbbgky6}).} If several minima exist at the same temperature,
the selection depends on a notion of basin of attraction.  The
relaxation time is $t_{B}\sim 1/\xi$, independent of $N$.

The mean field Smoluchowski equation (\ref{smf10g}) may also be written as
\begin{eqnarray}
{\partial\rho\over\partial
t}=\nabla\cdot \left\lbrack\frac{1}{\xi}\left (\nabla p+\rho \nabla\Phi\right )\right\rbrack, \label{smf10gbis}
\end{eqnarray}
where $p({\bf r},t)$ is a pressure related to the density by the isothermal equation of state
\begin{eqnarray}
p({\bf r},t)=\rho({\bf r},t)\frac{k_B T}{m}.
\label{piso}
\end{eqnarray}
This equation of state can be obtained from the expression of the local kinetic pressure $p({\bf r},t)=\frac{1}{d}\int f({\bf r},{\bf v},t) [{\bf v}-{\bf u}({\bf r},t)]^2\, d{\bf v}$ where ${\bf u}({\bf r},t)=(1/\rho)\int f{\bf v}\, d{\bf v}$ is the local velocity, combined with the expression (\ref{smf11}) of the distribution function valid in the strong friction limit. The steady states of the mean field Smoluchowski equation satisfy the equation
\begin{eqnarray}
\nabla p+\rho\nabla\Phi={\bf 0},
\label{smf16c}
\end{eqnarray}
which may be interpreted as a condition of hydrostatic equilibrium. A generalization of these results to other barotropic equations of state $p(\rho)$ is developed in \cite{gen,spa,cll,nfp,longshort}.   In that case, the free energy is given by
\begin{eqnarray}
\label{newsmol}
F[\rho]={1\over 2}\int\rho\Phi \, d{\bf r}+\int\rho\int^{\rho}\frac{p(\rho')}{{\rho'}^2}\, d{\bf r},
\end{eqnarray}
up to an additional constant.

{\it Remark:} at $T=0$, the free energy reduces to the potential energy $W=(1/2)\int\rho\Phi\, d{\bf r}$ and the $H$-theorem (\ref{smf13}) becomes
$\dot W=-\int (\rho/\xi)(\nabla\Phi)^2\, d{\bf r}\le 0$. In that case, the system relaxes towards the state of minimum potential energy.

\section{The general solution of the initial value problem using Green functions}
\label{sec_g}

The mean field Kramers equation writes
\begin{equation}
\frac{\partial f}{\partial t}+{\bf v}\cdot \frac{\partial f}{\partial {\bf r}}-\nabla\Phi\cdot \frac{\partial f}{\partial {\bf v}}=\xi\frac{\partial}{\partial {\bf v}}\cdot \left (\frac{k_B T}{m}\frac{\partial f}{\partial {\bf v}}+f{\bf v}\right ),
\label{g1}
\end{equation}
\begin{equation}
\Phi({\bf r},t)=\int u({\bf r}-{\bf r}')\rho({\bf r}',t)\, d{\bf r}'.
\label{g2}
\end{equation}
The spatially homogeneous steady state of this equation is the Maxwell-Boltzmann
distribution
\begin{equation}
f({\bf v})=\left (\frac{\beta m}{2\pi}\right )^{d/2}\rho\, e^{-\frac{1}{2}\beta m v^2}.
\label{g4}
\end{equation}
Considering a small perturbation $\delta f({\bf r},{\bf v},t)\ll f({\bf v})$ about this steady state, we obtain the linearized mean field Kramers equation
\begin{equation}
\frac{\partial \delta f}{\partial t}+{\bf v}\cdot \frac{\partial \delta f}{\partial {\bf r}}-\nabla\delta\Phi\cdot \frac{\partial f}{\partial {\bf v}}=\xi\frac{\partial}{\partial {\bf v}}\cdot \left (\frac{k_B T}{m}\frac{\partial \delta f}{\partial {\bf v}}+\delta f{\bf v}\right ),
\label{g1b}
\end{equation}
\begin{equation}
\delta \Phi({\bf r},t)=\int u({\bf r}-{\bf r}')\delta \rho({\bf r}',t)\, d{\bf r}'.
\label{g2b}
\end{equation}
This equation may be rewritten as
\begin{equation}
{\cal L}\delta f\equiv \frac{\partial \delta f}{\partial t}+{\bf v}\cdot \frac{\partial \delta f}{\partial {\bf r}}-\xi\frac{\partial}{\partial {\bf v}}\cdot \left (\frac{k_B T}{m}\frac{\partial \delta f}{\partial {\bf v}}+\delta f{\bf v}\right )=\frac{\partial f}{\partial {\bf v}}\cdot \int \nabla u({\bf r}-{\bf r}')\delta f({\bf r}',{\bf v}',t)\, d{\bf r}'d{\bf v}',
\label{g5}
\end{equation}
where ${\cal L}$ is the ordinary Kramers operator. To solve this equation we shall use the method of Green functions that has been introduced in similar problems \cite{gross}. The Green function of the ordinary Kramers operator is defined by
\begin{equation}
{\cal L}G({\bf r}-{\bf r}_0,{\bf v},{\bf v}_0,t)=\delta({\bf r}-{\bf r}_0)\delta({\bf v}-{\bf v}_0)\delta(t),
\label{g6}
\end{equation}
if $t\ge 0$ and $G({\bf r}-{\bf r}_0,{\bf v},{\bf v}_0,t)=0$ if
$t<0$. It depends only on the space variables ${\bf r}$ and ${\bf r}_0$ through the difference
${\bf x}={\bf r}-{\bf r}_0$. The solution of the initial value problem for the ordinary
Kramers equation is therefore
\begin{equation}
\delta f({\bf r},{\bf v},t)=\int G({\bf r}-{\bf r}_0,{\bf v},{\bf v}_0,t) \delta f({\bf r}_0,{\bf v}_0,0)\, d{\bf r}_0 d{\bf v}_0.
\label{g6b}
\end{equation}
The  Green function of the linearized mean field Kramers equation (\ref{g5}) is defined by
\begin{equation}
{\cal L}g({\bf r}-{\bf r}_0,{\bf v},{\bf v}_0,t)-\frac{\partial f}{\partial {\bf v}}\cdot \int \nabla u({\bf r}-{\bf r}')g({\bf r}'-{\bf r}_0,{\bf v}',{\bf v}_0,t)\, d{\bf r}'d{\bf v}'=\delta({\bf r}-{\bf r}_0)\delta({\bf v}-{\bf v}_0)\delta(t),
\label{g7}
\end{equation}
if $t\ge 0$ and $g({\bf r}-{\bf r}_0,{\bf v},{\bf v}_0,t)=0$ if $t<0$. It obeys the integral equation
\begin{equation}
g({\bf x},{\bf v},{\bf v}_0,t)=G({\bf x},{\bf v},{\bf v}_0,t)+\int G({\bf x}-{\bf x}',{\bf v},{\bf v}',t-t')\frac{\partial f}{\partial {\bf v}'}\cdot \nabla' u({\bf x}'-{\bf x}'')g({\bf x}'',{\bf v}'',{\bf v}_0,t')\, d{\bf x}''d{\bf v}''d{\bf x}'d{\bf v}'dt',
\label{g8}
\end{equation}
as may be checked by applying the
operator ${\cal L}$. In Eq. (\ref{g8}) we must have $t-t'\ge 0$ and
$t'\ge 0$ (otherwise the Green functions vanish) so that $0\le t'\le
t$. This integral equation can then be solved by applying the
convolution theorem. To that purpose, we introduce the Fourier-Laplace transform
\begin{equation}
\tilde{g}({\bf k},{\bf v},{\bf v}_0,\omega)=\int \frac{d{\bf r}}{(2\pi)^d}\int_{0}^{+\infty}dt\, e^{-i({\bf k}\cdot{\bf r}-\omega t)}g({\bf x},{\bf v},{\bf v}_0,t).
\label{g9}
\end{equation}
This expression for the Laplace transform is valid for ${\rm Im}(\omega)$ sufficiently large. For the remaining part of the complex $\omega$ plane, it is defined by an analytic continuation. The inverse transform is
\begin{equation}
g({\bf x},{\bf v},{\bf v}_0,t)=\int d{\bf k}\int_{\cal C}\frac{d\omega}{2\pi}\, e^{i({\bf k}\cdot{\bf r}-\omega t)}\tilde{g}({\bf k},{\bf v},{\bf v}_0,\omega),
\label{g10}
\end{equation}
where the Laplace contour ${\cal C}$ in the complex $\omega$ plane must pass above all poles of the integrand. Taking the Fourier-Laplace transform of Eq. (\ref{g8}) we get
\begin{equation}
\tilde{g}({\bf k},{\bf v},{\bf v}_0,\omega)=\tilde{G}({\bf k},{\bf v},{\bf v}_0,\omega)+i (2\pi)^{2d}\int \tilde{G}({\bf k},{\bf v},{\bf v}',\omega) {\bf k}\cdot\frac{\partial f}{\partial {\bf v}'} \hat{u}(k) \tilde{g}({\bf k},{\bf v}'',{\bf v}_0,\omega)\, d{\bf v}''d{\bf v}'.
\label{g11}
\end{equation}
Defining
\begin{equation}
\tilde{H}({\bf k},{\bf v},\omega)=i (2\pi)^{2d}\hat{u}(k)\int \tilde{G}({\bf k},{\bf v},{\bf v}',\omega) {\bf k}\cdot\frac{\partial f}{\partial {\bf v}'} \, d{\bf v}',
\label{g12}
\end{equation}
and
\begin{equation}
\tilde{q}({\bf k},{\bf v}_0,\omega)=\int  \tilde{g}({\bf k},{\bf v},{\bf v}_0,\omega)\, d{\bf v},
\label{g13}
\end{equation}
the foregoing equation may be rewritten as
\begin{equation}
\tilde{g}({\bf k},{\bf v},{\bf v}_0,\omega)=\tilde{G}({\bf k},{\bf v},{\bf v}_0,\omega)+\tilde{H}({\bf k},{\bf v},\omega)\tilde{q}({\bf k},{\bf v}_0,\omega).
\label{g14}
\end{equation}
Integrating over ${\bf v}$, we obtain
\begin{equation}
\tilde{q}({\bf k},{\bf v}_0,\omega)=\tilde{Q}({\bf k},{\bf v}_0,\omega)+\tilde{P}({\bf k},\omega)\tilde{q}({\bf k},{\bf v}_0,\omega),
\label{g15}
\end{equation}
where we have defined
\begin{equation}
\tilde{Q}({\bf k},{\bf v}_0,\omega)=\int \tilde{G}({\bf k},{\bf v},{\bf v}_0,\omega)  \, d{\bf v},
\label{g16}
\end{equation}
and
\begin{equation}
\tilde{P}({\bf k},\omega)=\int  \tilde{H}({\bf k},{\bf v},\omega)\, d{\bf v}.
\label{g17}
\end{equation}
Solving Eq. (\ref{g15}), we get
\begin{equation}
\tilde{q}({\bf k},{\bf v}_0,\omega)=\frac{\tilde{Q}({\bf k},{\bf v}_0,\omega)}{1-\tilde{P}({\bf k},\omega)}.
\label{g18}
\end{equation}
Substituting this expression in Eq. (\ref{g14}), we finally obtain
\begin{equation}
\tilde{g}({\bf k},{\bf v},{\bf v}_0,\omega)=\tilde{G}({\bf k},{\bf v},{\bf v}_0,\omega)+\frac{\tilde{H}({\bf k},{\bf v},\omega)\tilde{Q}({\bf k},{\bf v}_0,\omega)}{1-\tilde{P}({\bf k},\omega)}.
\label{g19}
\end{equation}
This is the resolvent, i.e. the Fourier-Laplace transform of the Green function. It connects $\delta \tilde{f}({\bf k},{\bf v},\omega)$ to the initial value. Indeed, the evolution of the perturbed distribution function is given by
\begin{equation}
\delta f({\bf r},{\bf v},t)=\int g({\bf r}-{\bf r}_0,{\bf v},{\bf v}_0,t) \delta f({\bf r}_0,{\bf v}_0,0)\, d{\bf r}_0 d{\bf v}_0.
\label{g20}
\end{equation}
Taking the Fourier-Laplace transform of this expression, we get
\begin{equation}
\delta \tilde{f}({\bf k},{\bf v},\omega)=(2\pi)^d\int \tilde{g}({\bf k},{\bf v},{\bf v}_0,\omega) \delta \hat{f}({\bf k},{\bf v}_0,0)\, d{\bf v}_0,
\label{g21}
\end{equation}
where $\delta \hat{f}({\bf k},{\bf v}_0,0)$ is the Fourier transform of the initial perturbed distribution function. Substituting Eq. (\ref{g19}) in Eq. (\ref{g21}), we obtain
\begin{equation}
\delta \tilde{f}({\bf k},{\bf v},\omega)=(2\pi)^d\int \tilde{G}({\bf k},{\bf v},{\bf v}_0,\omega) \delta \hat{f}({\bf k},{\bf v}_0,0)\, d{\bf v}_0+(2\pi)^{d}\frac{\tilde{H}({\bf k},{\bf v},\omega)}{1-\tilde{P}({\bf k},\omega)}\int \tilde{Q}({\bf k},{\bf v}_0,\omega)  \delta \hat{f}({\bf k},{\bf v}_0,0)\, d{\bf v}_0.
\label{g22}
\end{equation}
This is the solution of the initial value
problem in Fourier-Laplace space.  Integrating over the velocity, we find that the Fourier-Laplace transform of the perturbed density is given by
\begin{equation}
\delta \tilde{\rho}({\bf k},\omega)=(2\pi)^d \frac{1}{\epsilon({\bf k},\omega)}\int \tilde{Q}({\bf k},{\bf v}_0,\omega)  \delta \hat{f}({\bf k},{\bf v}_0,0)\, d{\bf v}_0,
\label{g23}
\end{equation}
where we have introduced the dielectric function
\begin{equation}
\epsilon({\bf k},\omega)=1-\tilde{P}({\bf k},\omega).
\label{g24}
\end{equation}
This expression shows that $\tilde{P}({\bf k},\omega)$ is the polarization function \cite{bt,lr}.
The perturbed density $\delta\tilde{\rho}({\bf k},\omega)$ given by
Eq. (\ref{g23}) appears as a product of two factors: a ``universal'' factor
 $\epsilon({\bf k},\omega)^{-1}$ and an integral involving the initial condition
$\delta\hat{f}({\bf k},{\bf v}_0,0)$. The first factor is due to
collective effects. As we shall explain below, this term can produce
damped, steady, or growing oscillations. On the other hand, the integral corresponds to
the excess density produced by an initial disturbance in a Brownian
gas of non-interacting particles (i.e., for which $\hat{u}(k)=0$ or $\epsilon({\bf
k},\omega)=1$). It is typical of an individual particle behavior. 
The effect of this term disappears for late
times since the usual Kramers equation relaxes towards the Maxwell-Boltzmann distribution (\ref{g4}). 
For dissipationless systems ($\xi=0$) this term is responsible for the
phenomenon of ``phase mixing'' associated with the Vlasov equation even in the
absence of interaction (see Sec. \ref{sec_v}). 

The temporal evolution of the Fourier modes of the density perturbation is
given by the inverse Laplace transform 
\begin{equation}
\label{pr15}
\delta\hat{\rho}({\bf k},t)=\int_{\cal C}\frac{d\omega}{2\pi}\, e^{-i\omega t}\delta\tilde{\rho}({\bf k},\omega).
\end{equation}
The ``universal'' poles of $\delta\tilde{\rho}({\bf k},\omega)$ correspond
to the complex pulsations $\omega_{\alpha}({\bf k})$ for which the
dielectric function vanishes: $\epsilon({\bf k},\omega_{\alpha}({\bf
k}))=0$. This defines the dispersion relation. The evolution of the
perturbation depends on the position of the zeros of the dielectric
function in the complex plane. Using the Cauchy residue theorem, we
have
\begin{equation}
\label{pr16}
\delta\hat{\rho}({\bf k},t)=-i\sum_{\alpha} e^{-i\omega_{\alpha}({\bf k}) t} \left\lbrack {\rm Res} \, \delta\tilde{\rho}({\bf k},\omega) \right \rbrack_{\omega=\omega_{\alpha}({\bf k})},
\end{equation}
where the sum runs over the whole set of poles and we have assumed, for simplicity, that the singularities are simple poles. In the following, we shall omit the subscript $\alpha$ for brevity. If at least one zero $\omega$ of the dielectric function lies on the upper half plane (i.e. $\omega_i>0$), the system is unstable, and the perturbation grows exponentially rapidly with the rate $(\omega_i)_{max}$ corresponding to the zero with the largest value of the imaginary pulsation. If all the zeros $\omega$ of the dielectric function strictly lie on the lower half-plane (i.e. $\omega_i<0$), the system is stable, and the perturbation decays to zero exponentially rapidly with the rate $|\omega_i|_{min}$ corresponding to the zero with the smallest value of the imaginary pulsation in absolute value. If some zero(s) lie(s) on the real axis (i.e. $\omega_i=0$) while the others lie on the lower half-plane, the system is marginally stable and the perturbation displays an oscillating behavior around zero with the pulsation(s) $\omega_r$. If the integrand has a pole at $\omega=0$ while the other zeros lie on the lower half-plane, the perturbation tends to a steady state for $t\rightarrow +\infty$. Finally, if the integrand has a pole at $\omega=0$ while other zeros lie on the real axis and the rest on the lower half-plane, the perturbation oscillates about a steady state. For more details, we refer to \cite{balescubook,bt}.

{\it Remark:} Although we have considered the Kramers operator for illustration, we
emphasize that the results of this section are actually valid for any
linear operator ${\cal L}$. Indeed, the formal solution of the problem
only involves the Green function of the operator ${\cal L}$ and the
steady distribution $f({\bf v})$. In this sense, the preceding formalism is very general.

\section{Initial value problem for the linearized Vlasov equation}
\label{sec_v}

\subsection{The dielectric function}

If we take  $\xi=0$ in Eq. (\ref{g1}), we obtain the Vlasov
equation. In that case, we can consider any steady state of the form
$f=f({\bf v})$, not only the Maxwellian. Let us check that the general
formalism developed previously returns the classical results for the
initial value problem of the linearized Vlasov equation. The Green
function of a free particle ($\xi=0$) is simply
\begin{equation}
G({\bf r}-{\bf r}_0,{\bf v},{\bf v}_0,t)=\delta({\bf v}-{\bf v}_0)\delta({\bf r}-{\bf r}_0-{\bf v}_0t).
\label{v1}
\end{equation}
Its Fourier transform is
\begin{equation}
\hat{G}({\bf k},{\bf v},{\bf v}_0,t)=\frac{1}{(2\pi)^d}\delta({\bf v}-{\bf v}_0)e^{-i {\bf k}\cdot {\bf v}_0 t},
\label{v1b}
\end{equation}
and its Fourier-Laplace transform is
\begin{equation}
\tilde{G}({\bf k},{\bf v},{\bf v}_0,\omega)=i\frac{1}{(2\pi)^d}\delta({\bf v}-{\bf v}_0)\frac{1}{\omega-{\bf k}\cdot {\bf v}_0}.
\label{v2}
\end{equation}
From this expression, we obtain
\begin{equation}
\tilde{H}({\bf k},{\bf v},\omega)=(2\pi)^d \hat{u}(k)\frac{{\bf k}\cdot \frac{\partial f}{\partial {\bf v}}}{{\bf k}\cdot {\bf v}-\omega},\qquad \tilde{Q}({\bf k},{\bf v}_0,\omega)=i\frac{1}{(2\pi)^d}\frac{1}{\omega-{\bf k}\cdot {\bf v}_0},
\label{v3}
\end{equation}
\begin{equation}
\tilde{P}({\bf k},\omega)=(2\pi)^d \hat{u}(k)\int \frac{{\bf k}\cdot \frac{\partial f}{\partial {\bf v}}}{{\bf k}\cdot {\bf v}-\omega}\, d{\bf v},\qquad \epsilon({\bf k},\omega)=1-(2\pi)^d\hat{u}(k)\int \frac{{\bf k}\cdot \frac{\partial f}{\partial {\bf v}}}{{\bf k}\cdot {\bf v}-\omega}\, d{\bf v}.
\label{v4}
\end{equation}
Using Eqs. (\ref{g12}), (\ref{g17}) and (\ref{v1b}) we find that the temporal evolution of the polarization function is (see also \cite{lr}):
\begin{equation}
\hat{P}({\bf k},t)=-(2\pi)^d \hat{u}(k)\rho k^2 t e^{-\frac{k^2 t^2}{2\beta m}}.
\label{v4b}
\end{equation}
According to Eq. (\ref{g22}), the solution of the initial value problem is
\begin{equation}
\delta\tilde{f}({\bf k},{\bf v},\omega)=i\frac{\delta\hat{f}({\bf k},{\bf v},0)}{\omega-{\bf k}\cdot {\bf v}}+i\frac{(2\pi)^d\hat{u}(k)}{\epsilon({\bf k},\omega)}\frac{{\bf k}\cdot \frac{\partial f}{\partial {\bf v}}}{{\bf k}\cdot {\bf v}-\omega}\int \frac{\delta\hat{f}({\bf k},{\bf v}_0,0)}{\omega-{\bf k}\cdot {\bf v}_0}\, d{\bf v}_0.
\label{v7}
\end{equation}
The resolvent operator that connects $\delta\tilde{f}({\bf
k},{\bf v},\omega)$ to the initial value through Eq. (\ref{g21}) is
\begin{equation}
\label{iv5}
\tilde{g}({\bf k},{\bf v},{\bf v}_0,\omega)=\frac{1}{(2\pi)^d}\frac{\delta({\bf v}-{\bf v}_0)}{i({\bf k}\cdot {\bf v}-\omega)}+\frac{{\bf k}\cdot \frac{\partial f}{\partial {\bf v}}}{{\bf k}\cdot {\bf v}-\omega}\frac{\hat{u}(k)}{\epsilon({\bf k},\omega)}\frac{1}{i({\bf k}\cdot {\bf v}_0-\omega)}.
\end{equation}
The density perturbation is given by
\begin{equation}
\delta\tilde{\rho}({\bf k},\omega)=i\frac{1}{\epsilon({\bf k},\omega)}\int \frac{\delta\hat{f}({\bf k},{\bf v}_0,0)}{\omega-{\bf k}\cdot {\bf v}_0}\, d{\bf v}_0.
\label{v8}
\end{equation}
This returns the results obtained by directly taking the Fourier-Laplace transform of the linearized Vlasov equation corresponding to Eqs. (\ref{g1b}) and (\ref{g2b}) with $\xi=0$. Indeed, they give
\begin{equation}
\label{iv1}
\delta\tilde{f}({\bf k},{\bf v},\omega)=\frac{{\bf k}\cdot \frac{\partial f}{\partial {\bf v}}}{{\bf k}\cdot {\bf v}-\omega} \delta\tilde\Phi({\bf k},\omega)+\frac{\delta\hat{f}({\bf k},{\bf v},0)}{i({\bf k}\cdot {\bf v}-\omega)},
\end{equation}
\begin{equation}
\label{vl4}
\delta\tilde\Phi({\bf k},\omega)=(2\pi)^d\hat{u}(k)\delta \tilde\rho({\bf k},\omega).
\end{equation}
Integrating Eq. (\ref{iv1}) over the velocity and using
Eq. (\ref{vl4}), we get Eq. (\ref{v8}). Substituting this result back
into Eq. (\ref{iv1}), we recover Eq. (\ref{v7}).

The integral in Eq. (\ref{v8}) corresponds to the excess density produced by an
initial disturbance in a gas of non-interacting particles. The effect of this
term disappears for late times. Indeed, it can be shown that this integral
produces damped oscillations (i.e. its poles are in the lower half-plane).
Therefore, the density perturbation $\delta\hat{\rho}({\bf k},t)$ decays to zero
although the Vlasov equation is time reversible. By contrast, the perturbed
distribution function (\ref{v7}) has an additional real pole $\omega={\bf
k}\cdot {\bf v}$. It produces an undamped oscillation ${\rm exp}(-i{\bf k}\cdot
{\bf v}t)$ whose pulsation is proportional to the velocity ${\bf v}$ of the
particles. Therefore the distribution function does not decay to zero but
generates small-scale filaments. However, if we consider the perturbed density
$\delta\hat{\rho}({\bf k},t)=\int \delta\hat{f}({\bf k},{\bf v},t)\, d{\bf v}$
obtained by integrating the perturbed distribution function $\delta\hat{f}({\bf
k},{\bf v},t)$ over the velocity, the various velocities produce destructive
interferences of the oscillations,  and this is why the density perturbation
decays: this is the phenomenon of phase mixing. This is an irreversible
homogenization process in which the interactions play no role. The other poles
of Eq. (\ref{v7}) correspond to the zeros of the dielectric function. They
depend on ${\bf k}$ but not on ${\bf v}$. They describe the collective behavior
of the system. They produce damped or growing oscillations that ``resist'' the
integration over ${\bf v}$ \cite{balescubook}.

\subsection{The dispersion relation and the stability criterion}

Although we can study the dispersion relation of the linearized Vlasov
equation for any steady distribution $f({\bf v})$, we restrict
ourselves here to the the case of the Maxwellian because we ultimately want to
compare the results obtained from the linearized Vlasov equation to the results
obtained from the linearized mean field Kramers equation that are valid only for
the Maxwellian. Other steady states of the Vlasov equation are considered in
\cite{cd,nyquistgrav,lr} and in classical textbooks of
plasma physics \cite{balescubook}.

For the Maxwell-Boltzmann distribution (\ref{g4}), we can write the dielectric function in the form
\begin{equation}
\label{v9}
\epsilon(k,\omega)=1+(2\pi)^d\hat{u}(k)\rho\beta m  W\left (\sqrt{\beta m}\frac{\omega}{k}\right),
\end{equation}
where
\begin{equation}
\label{v10}
W(z)=\frac{1}{\sqrt{2\pi}}\int_{L}\frac{x}{x-z}e^{-x^2/2}\, dx,
\end{equation}
is the plasma dispersion function \cite{fried}. The integration has to be performed along the Landau contour $L$ \cite{landau}.  For any complex $z$, we have
\begin{equation}
\label{v11}
W(z)=1-ze^{-z^2/2}\int_0^z e^{y^2/2}\, dy+i\sqrt{\frac{\pi}{2}}z e^{-z^2/2}.
\end{equation}
The dispersion relation $\epsilon({k},\omega)=0$ can be written as
\begin{equation}
1+(2\pi)^d\hat{u}(k)\rho\beta m  W\left (\sqrt{\beta m}\frac{\omega}{k}\right)=0.
\label{v12}
\end{equation}
The neutral mode corresponds to $\omega=0$. Using $W(0)=1$ we get the condition $1+(2\pi)^d\hat{u}(k)\rho\beta m=0$. Using the Nyquist theorem \cite{balescubook,cd,nyquistgrav}, we can show that the system is stable with respect to a perturbation of wavenumber $k$ when
\begin{equation}
1+(2\pi)^d\hat{u}(k)\rho\beta m>0,
\label{v14}
\end{equation}
and unstable otherwise. For repulsive potentials for which $\hat{u}(k)>0$, the system is always stable. For attractive potentials for which $\hat{u}(k)<0$, the system is always stable when $T>T_c=(\rho m/k_B)(2\pi)^d\max_k|\hat{u}(k)|$ while it is unstable to some modes (corresponding to the converse of Eq. (\ref{v14})) when $T<T_c$. This stability criterion can also be obtained from the condition of formal nonlinear dynamical stability (see Appendix \ref{sec_so}).

We look for solutions of the dispersion relation (\ref{v12}) in the
form $\omega=i\omega_i$ where $\omega_i$ is real. When $\omega_i>0$, the perturbation grows exponentially rapidly and when $\omega_i<0$ is decays exponentially rapidly (without oscillating). The growth or decay rate $\omega_i$ is given by
\begin{equation}
\label{v15}
1+(2\pi)^d\hat{u}(k)\rho\beta m  w\left (\sqrt{\frac{\beta m}{2}}\frac{\omega_i}{k}\right)=0,
\end{equation}
with
\begin{equation}
\label{v16}
w(x)=1-\sqrt{\pi}x e^{x^2}{\rm erfc}(x),
\end{equation}
where erfc is the complementary error function defined by
\begin{equation}
\label{v17}
{\rm erfc}(x)=1-{\rm erf}(x),\qquad {\rm erf}(x)=\frac{2}{\sqrt{\pi}}\int_0^x e^{-y^2}\, dy.
\end{equation}
We note that $W(i x)=w(x/\sqrt{2})$ for any real $x$.   This function has the asymptotic behaviors
\begin{equation}
\label{v18}
w(x)\simeq 1-\sqrt{\pi}x \qquad (x\rightarrow 0),
\end{equation}
\begin{equation}
\label{v19}
w(x)\sim \frac{1}{2x^2}\left (1-\frac{3}{2x^2}\right ) \qquad (x\rightarrow +\infty),
\end{equation}
\begin{equation}
\label{v20}
w(x)\sim 2\sqrt{\pi}|x|e^{x^2} \qquad (x\rightarrow -\infty).
\end{equation}
The asymptotic behaviors of the inverse function are
\begin{equation}
\label{v21}
w^{-1}(y)\sim \frac{1}{\sqrt{\pi}}(1-y) \qquad (y\rightarrow 1),
\end{equation}
\begin{equation}
\label{v22}
w^{-1}(y)\sim \frac{1}{\sqrt{2y}}\left (1-\frac{3}{2}y\right ) \qquad (y\rightarrow 0),
\end{equation}
\begin{equation}
\label{v23}
w^{-1}(y)\sim -(\ln y)^{1/2} \qquad (y\rightarrow +\infty).
\end{equation}
In the unstable case  ($\omega_i>0$), using the Nyquist theorem \cite{balescubook,cd,nyquistgrav},
we can show that the purely imaginary pulsation $\omega=i\omega_i$
determined by Eq. (\ref{v15}) is the only solution of the dispersion
relation (\ref{v12}). In the stable case ($\omega_i<0$), there exist other solutions
of the form $\omega=\omega_r+i\omega_i$ with $\omega_r\neq 0$.

\subsection{Application to the HMF model, self-gravitating systems, and plasmas}

For illustration, we apply the preceding results to the attractive and repulsive HMF models, self-gravitating systems, and plasmas. For the definition of these models and for the notations we refer to \cite{cd,nyquistgrav,lr}.

For the attractive HMF model, using $\hat{u}_n=\frac{1}{2N}(2\delta_{n,0}-\delta_{n,1}-\delta_{n,-1})$ and $\rho=1/(2\pi)$, and considering the modes $n=\pm 1$ (the modes $n\neq \pm 1$ cannot propagate), the dispersion relation (\ref{v12}) can be written as
\begin{equation}
1-\frac{1}{2T} W\left (\frac{\omega}{\sqrt{T}}\right)=0.
\label{v29}
\end{equation}
According to Eq. (\ref{v14}) the system is stable if $T>T_c=1/2$ and unstable with respect to the modes $n=\pm 1$ if $T<T_c$. Assuming that $\omega=i\omega_i$, we get
\begin{equation}
\omega_i=\sqrt{2T} w^{-1}(2T).
\label{v30}
\end{equation}
We have the asymptotic behaviors
\begin{equation}
{\omega_i}\simeq \frac{1}{\sqrt{2}}(1-3T) \qquad (T\rightarrow 0),
\label{v31}
\end{equation}
\begin{equation}
{\omega_i}\simeq \frac{1}{\sqrt{\pi}}(1-2T) \qquad (T\rightarrow T_c),
\label{v32}
\end{equation}
\begin{equation}
{\omega_i}\sim -\sqrt{2T\ln T} \qquad (T\rightarrow +\infty).
\label{v33}
\end{equation}

For self-gravitating systems, using $(2\pi)^d \hat{u}(k)=-S_d G/k^2$
and making the Jeans swindle (see \cite{bt,nyquistgrav} for more
details), the dispersion relation (\ref{v12}) can be written as
\begin{equation}
1-\frac{k_J^2}{k^2} W\left (\frac{\omega}{\omega_G}\frac{k_J}{k}\right)=0,
\label{v24}
\end{equation}
where we have introduced the Jeans wavenumber $k_J=(S_d G\rho\beta m)^{1/2}$ and the gravitational pulsation $\omega_G=(S_d G\rho)^{1/2}$ (the inverse of the dynamical time $t_D=1/\omega_G$). According to Eq. (\ref{v14}) the system is stable if $k>k_J$ and unstable if $k<k_J$. Assuming that $\omega=i\omega_i$, we get
\begin{equation}
\frac{\omega_i}{\omega_G}=\sqrt{2}\frac{k}{k_J} w^{-1}\left (\frac{k^2}{k_J^2}\right ).
\label{v25}
\end{equation}
We have the asymptotic behaviors
\begin{equation}
\frac{\omega_i}{\omega_G}\simeq 1-\frac{3}{2}\frac{k^2}{k_J^2} \qquad (k\rightarrow 0),
\label{v26}
\end{equation}
\begin{equation}
\frac{\omega_i}{\omega_G}\simeq \sqrt{\frac{2}{\pi}}\left (1-\frac{k^2}{k_J^2}\right ) \qquad (k\rightarrow k_J),
\label{v27}
\end{equation}
\begin{equation}
\frac{\omega_i}{\omega_G}\sim -2\frac{k}{k_J}\sqrt{\ln\left (\frac{k}{k_J}\right )} \qquad (k\rightarrow +\infty).
\label{v28}
\end{equation}

For the repulsive HMF model, using $\hat{u}_n=-\frac{1}{2N}(2\delta_{n,0}-\delta_{n,1}-\delta_{n,-1})$ and $\rho=1/(2\pi)$, and considering the modes $n=\pm 1$ (the modes $n\neq \pm 1$ cannot propagate), the dispersion relation (\ref{v12}) can be written as
\begin{equation}
1+\frac{1}{2T} W\left (\frac{\omega}{\sqrt{T}}\right)=0.
\label{v37}
\end{equation}
According to Eq. (\ref{v14}), the system is always stable. There is no solution 
of the dispersion
relation (\ref{v37}) of the form $\omega=i\omega_i$. However, some asymptotic
solutions of Eq. (\ref{v37}) can be obtained \cite{cd}. In the limit
$T\rightarrow 0$  we have $\omega=\omega_r+i\omega_i$ with
$\omega_i\ll\omega_r$ and the solution of the dispersion relation is 
\begin{equation}
\omega_r^2\simeq \frac{1}{2}+3T+...,\qquad \omega_i\sim 
-\frac{1}{8}\sqrt{\frac{\pi}{2}}\frac{1}{T^{3/2}}e^{-\frac{1}{4T}}\qquad
(T\rightarrow 0).
\label{v38}
\end{equation}
At $T=0$, the perturbation oscillates with the pulsation $\omega_r=1/\sqrt{2}$. For $T>0$, it also experiences a weak Landau damping $\omega_i<0$. In the limit $T\rightarrow +\infty$ we have $\omega=\omega_r+i\omega_i$ with $\omega_i\gg\omega_r$ and the solution of the dispersion relation is
\begin{equation}
\omega_r\sim \pi\sqrt{\frac{T}{2\ln T}}, \qquad \omega_i\sim -\sqrt{2T\ln T}
\qquad (T\rightarrow +\infty).
\label{v39}
\end{equation}
In that case, the perturbation exhibits heavily damped oscillations.

For Coulombian plasmas, using $(2\pi)^d \hat{u}(k)=S_d e^2/m^2k^2$, the dispersion relation (\ref{v12}) can be written as
\begin{equation}
1+\frac{k_D^2}{k^2} W\left (\frac{\omega}{\omega_P}\frac{k_D}{k}\right)=0,
\label{v34}
\end{equation}
where we have introduced the Debye wavenumber $k_D=(S_d \rho
e^2\beta/m)^{1/2}$ and the plasma pulsation $\omega_P=(S_d \rho e^2/m^2)^{1/2}$
(the inverse of the dynamical time $t_D=1/\omega_P$). According to Eq.
(\ref{v14}) the system is always stable. There is no solution
of the dispersion relation (\ref{v34}) of the form $\omega=i\omega_i$. However,
some asymptotic solutions of Eq. (\ref{v34}) can be obtained \cite{balescubook}.
For $k\ll k_D$ (long wavelengths) we have $\omega=\omega_r+i\omega_i$ with
$\omega_i\ll\omega_r$ and the solution of the dispersion relation is
\begin{equation}
\omega_r^2\simeq \omega_P^2+3\frac{k_B T}{m}k^2+...,\qquad 
\omega_i\sim -\sqrt{\frac{\pi}{8}}\omega_P\left (\frac{k_D}{k}\right )^3
e^{-\frac{k_D^2}{2k^2}}\qquad (k\rightarrow 0).
\label{v35}
\end{equation}
For $k=0$, the perturbation oscillates with the plasma pulsation $\omega_P$. For
$k>0$, it also 
experiences a weak Landau damping \cite{landau}. For $k\gg k_D$ (small
wavelengths) we have $\omega=\omega_r+i\omega_i$ with $\omega_i\gg\omega_r$ and
the solution of the dispersion relation is
\begin{equation}
\omega_r\sim \frac{\pi}{2}\omega_P \frac{k}{k_D}\frac{1}{\sqrt{\ln(k/k_D)}}, 
\qquad \omega_i\sim -2\omega_P\frac{k}{k_D} \sqrt{\ln(k/k_D)} \qquad
(k\rightarrow +\infty).
\label{v36}
\end{equation}
In that case, the perturbation exhibits heavily damped oscillations.

\section{Initial value problem for the linearized mean field Smoluchowski equation}
\label{sec_s}

\subsection{The dielectric function by a direct approach}

The mean field Smoluchowski equation writes
\begin{eqnarray}
{\partial\rho\over\partial
t}=\frac{1}{\xi}\nabla\cdot (\nabla p+\rho\nabla\Phi), \label{s1}
\end{eqnarray}
\begin{equation}
\Phi({\bf r},t)=\int u({\bf r}-{\bf r}')\rho({\bf r}',t)\, d{\bf r}'.
\label{s2}
\end{equation}
For the sake of generality, we consider an arbitrary barotropic
equation of state $p=p(\rho)$. This leads to the generalized mean field
Smoluchowski equation \cite{gen,spa,cll,nfp,longshort}.  For example, the polytropic
equation of state $p=K\rho^{\gamma}$ can account for anomalous diffusion like
in porous media. The
usual Smoluchowski equation corresponds to the isothermal equation of
state (\ref{piso}) leading to normal diffusion with the diffusion
coefficient (\ref{einstein2}).

Considering  a small perturbation $\delta\rho({\bf r},t)\ll \rho$ about a spatially homogeneous steady state, we obtain the linearized mean field Smoluchowski equation
\begin{eqnarray}
{\partial\delta\rho\over\partial
t}=\frac{1}{\xi}\nabla\cdot (c_s^2\nabla \delta\rho+\rho\nabla\delta\Phi), \label{s3}
\end{eqnarray}
\begin{equation}
\delta\Phi({\bf r},t)=\int u({\bf r}-{\bf r}')\delta\rho({\bf r}',t)\, d{\bf r}',
\label{s4}
\end{equation}
where $c_s^2=p'(\rho)$ is the velocity of sound. Taking the Fourier-Laplace transform of these equations, we obtain
\begin{eqnarray}
-\xi\delta\hat{\rho}({\bf k},0)-i\xi\omega\delta\tilde{\rho}({\bf k},\omega)=-c_s^2 k^2\delta\tilde{\rho}({\bf k},\omega)-\rho k^2\delta\tilde{\Phi}({\bf k},\omega),
\label{s5}
\end{eqnarray}
\begin{equation}
\delta\tilde{\Phi}({\bf k},\omega)=(2\pi)^d \hat{u}(k)\delta\tilde{\rho}({\bf k},\omega),
\label{s6}
\end{equation}
where $\delta\hat{\rho}({\bf k},0)$ is the Fourier transform of the initial perturbation $\delta{\rho}({\bf r},0)$. Solving these equations, we get
\begin{eqnarray}
\delta\tilde{\rho}({\bf k},\omega)=\frac{\xi\delta\hat{\rho}({\bf k},0)}{-i\xi\omega+c_s^2 k^2+(2\pi)^d\hat{u}(k)\rho k^2}.
\label{s7}
\end{eqnarray}
This is the exact solution of the initial value problem for the linearized Smoluchowski equation in Fourier-Laplace space. Equation (\ref{s7}) can be rewritten as
\begin{eqnarray}
\delta\tilde{\rho}({\bf k},\omega)= \frac{1}{\epsilon(k,\omega)}\frac{\delta\hat{\rho}({\bf k},0)}{\frac{c_s^2}{\xi}k^2-i\omega},
\label{s8}
\end{eqnarray}where we have introduced the dielectric function \cite{lr}:
\begin{eqnarray}
\epsilon({k},\omega)=1-\tilde{P}(k,\omega)=1-\frac{(2\pi)^d \hat{u}(k)\rho k^2}{i\xi\omega-c_s^2 k^2}.
\label{s9}
\end{eqnarray}
Taking the inverse Laplace transform of Eq. (\ref{s7}) and using the Cauchy residue theorem, we find that the temporal evolution of the Fourier components of the density perturbation is
\begin{eqnarray}
\delta\hat{\rho}({\bf k},t)=\delta\hat{\rho}({\bf k},0) e^{-\left\lbrack c_s^2+(2\pi)^d \hat{u}(k)\rho\right\rbrack k^2 t/\xi}.
\label{s9b}
\end{eqnarray}
Actually, this result may be directly obtained by taking the Fourier transform of Eqs. (\ref{s3}) and (\ref{s4}) which leads to the first order equation in time
\begin{eqnarray}
\frac{d\delta\hat{\rho}}{dt}+\frac{1}{\xi}\left\lbrack c_s^2+(2\pi)^d \hat{u}(k)\rho\right\rbrack k^2 \delta\hat{\rho}=0.
\label{s9c}
\end{eqnarray}
Integrating this equation, we obtain Eq. (\ref{s9b}).

\subsection{The dielectric function by using the Green function}

It is instructive to recover these results by using the same method as
in Sec. \ref{sec_g}. The linearized mean field Smoluchowski equation
may be rewritten as
\begin{equation}
{\cal L}\delta \rho\equiv \frac{\partial \delta \rho}{\partial t}-D_*\Delta\delta\rho=\frac{1}{\xi}\rho\int \Delta u({\bf r}-{\bf r}')\delta \rho({\bf r}',t)\, d{\bf r}',
\label{s10}
\end{equation}
where ${\cal L}$ is the ordinary diffusion operator with diffusion coefficient $D_*=c_s^2/\xi$. The Green function of the ordinary diffusion operator is defined by
\begin{equation}
{\cal L}G({\bf r}-{\bf r}_0,t)=\delta({\bf r}-{\bf r}_0)\delta(t),
\label{s11}
\end{equation}
if $t\ge 0$ and $G({\bf r}-{\bf r}_0,t)=0$ if
$t<0$. It depends only on the space variables ${\bf r}$ and ${\bf r}_0$ through the difference
${\bf x}={\bf r}-{\bf r}_0$. The solution of the initial value problem for the ordinary
diffusion equation is therefore
\begin{equation}
\delta \rho({\bf r},t)=\int G({\bf r}-{\bf r}_0,t) \delta \rho({\bf r}_0,0)\, d{\bf r}_0 .
\label{s12}
\end{equation}
The  Green function of the linearized mean field Smoluchowski equation is defined by
\begin{equation}
{\cal L}g({\bf r}-{\bf r}_0,t)-\frac{1}{\xi}\rho \int \Delta u({\bf r}-{\bf r}')g({\bf r}'-{\bf r}_0,t)\, d{\bf r}'=\delta({\bf r}-{\bf r}_0)\delta(t),
\label{s13}
\end{equation}
if $t\ge 0$ and $g({\bf r}-{\bf r}_0,t)=0$ if $t<0$. It obeys the integral equation
\begin{equation}
g({\bf x},t)=G({\bf x},t)+\frac{1}{\xi}\rho \int G({\bf x}-{\bf x}',t-t')\Delta' u({\bf x}'-{\bf x}'')g({\bf x}'',t')\, d{\bf x}''d{\bf x}'dt',
\label{s14}
\end{equation}
as may be checked by applying the
operator ${\cal L}$. In Eq. (\ref{s14}) we must have $t-t'\ge 0$ and
$t'\ge 0$ (otherwise the Green functions vanish) so that $0\le t'\le
t$. This integral equation can then be solved by applying the
convolution theorem. Taking the Fourier-Laplace transform of Eq. (\ref{s14}) we get
\begin{equation}
\tilde{g}({\bf k},\omega)=\frac{\tilde{G}({\bf k},\omega)}{1+\frac{1}{\xi}\rho (2\pi)^{2d} \hat{u}(k) k^2 \tilde{G}({\bf k},\omega)}.
\label{s15}
\end{equation}
This is the resolvent operator, i.e. the Fourier-Laplace transform of the Green function. It connects $\delta \tilde{\rho}({\bf k},\omega)$ to the initial value. Indeed, the evolution of the perturbed density is given by
\begin{equation}
\delta \rho({\bf r},t)=\int g({\bf r}-{\bf r}_0,t) \delta \rho({\bf r}_0,0)\, d{\bf r}_0.
\label{s16}
\end{equation}
Taking the Fourier-Laplace transform of this expression, we get
\begin{equation}
\delta \tilde{\rho}({\bf k},\omega)=(2\pi)^d\tilde{g}({\bf k},\omega) \delta \hat{\rho}({\bf k},0).
\label{s17}
\end{equation}
Substituting Eq. (\ref{s15}) in Eq. (\ref{s17}), we obtain
\begin{equation}
\delta \tilde{\rho}({\bf k},\omega)=(2\pi)^d \frac{1}{\epsilon({\bf k},\omega)}\tilde{G}({\bf k},\omega)  \delta \hat{\rho}({\bf k},0),
\label{s18}
\end{equation}
where we have introduced the dielectric function
\begin{equation}
\epsilon({\bf k},\omega)=1-\tilde{P}({\bf k},\omega)=1+\frac{1}{\xi}\rho (2\pi)^{2d}\hat{u}(k)k^2\tilde{G}({\bf k},\omega).
\label{s19}
\end{equation}
The Green function of the ordinary diffusion equation is
\begin{equation}
G({\bf r}-{\bf r}_0,t)=\frac{1}{(4\pi D_*t)^{d/2}}e^{-\frac{|{\bf r}-{\bf r}_0|^2}{4D_*t}}.
\label{s20}
\end{equation}
Its Fourier transform is
\begin{equation}
\hat{G}({\bf k},t)=\frac{1}{(2\pi)^d}e^{-D_* t k^2},
\label{s21}
\end{equation}
and its Fourier-Laplace transform is
\begin{equation}
\tilde{G}({\bf k},\omega)=\frac{1}{(2\pi)^d}\frac{1}{D_* k^2-i\omega}.
\label{s22}
\end{equation}
Substituting Eq. (\ref{s22}) in Eqs. (\ref{s18}) and (\ref{s19}),
we recover Eq. (\ref{s8}). From Eqs. (\ref{s19}) and (\ref{s21}) the temporal evolution of the polarization function is given by
\begin{equation}
\hat{P}({\bf k},t)=-\frac{1}{\xi} (2\pi)^{d}\hat{u}(k)\rho k^2 e^{-c_s^2k^2t/\xi}.
\label{s22b}
\end{equation}

\subsection{The dispersion relation and the stability criterion}

The dispersion relation $\epsilon({k},\omega)=0$ can be written as
\begin{equation}
i\xi\omega=c_s^2k^2+(2\pi)^d\hat{u}(k)\rho k^2.
\label{s23}
\end{equation}
The complex pulsation is purely imaginary: $\omega=i\omega_i$. The perturbation grows exponentially rapidly when $\omega_i>0$ and it decays exponentially rapidly when $\omega_i<0$ (without oscillating). The neutral mode corresponds to $\omega=0$. We get the condition $c_s^2+(2\pi)^d\hat{u}(k)\rho=0$.
The system is stable with respect to a perturbation with wavenumber $k$ when
\begin{equation}
c_s^2+(2\pi)^d\hat{u}(k)\rho>0,
\label{s25}
\end{equation}
and unstable otherwise. This
stability criterion can also be obtained from the study of the second
order variations of the free energy (see Appendix \ref{sec_so}).

\subsection{Application to the BMF model, self-gravitating systems, and plasmas}

For the attractive BMF model, using $\hat{u}_n=\frac{1}{2N}(2\delta_{n,0}-\delta_{n,1}-\delta_{n,-1})$ and $\rho=1/(2\pi)$,  the dispersion relation (\ref{s23}) can be written as
\begin{equation}
i\xi\omega=c_s^2n^2,\qquad (n\neq \pm 1), \qquad i\xi\omega=c_s^2-\frac{1}{2}, \qquad (n= \pm 1).
\label{s27}
\end{equation}
The modes $n\neq \pm 1$ are damped exponentially rapidly (stable). The modes $n=\pm 1$ are damped  exponentially rapidly if $c_s^2>1/2$ (stable) and they grow exponentially rapidly if $c_s^2<1/2$ (unstable).

For self-gravitating systems, using $(2\pi)^d \hat{u}(k)=-S_d G/k^2$, the dispersion relation (\ref{s23}) can be written as
\begin{equation}
i\xi\omega=c_s^2k^2-S_d G\rho.
\label{s26}
\end{equation}
The system is stable if $k>k_J$ and unstable if $k<k_J$.

For the repulsive BMF model, using
$\hat{u}_n=-\frac{1}{2N}(2\delta_{n,0}-\delta_{n,1}-\delta_{n,-1})$
and $\rho=1/(2\pi)$, the dispersion relation (\ref{s23}) can be written as
\begin{equation}
i\xi\omega=c_s^2n^2,\qquad (n\neq \pm 1),\qquad i\xi\omega=c_s^2+\frac{1}{2}, \qquad (n= \pm 1).
\label{s30}
\end{equation}
The system is always stable.

For Coulombian plasmas, using $(2\pi)^d \hat{u}(k)=S_d e^2/m^2k^2$, the dispersion relation (\ref{s23}) can be written as
\begin{equation}
i\xi\omega=c_s^2k^2+\omega_P^2.
\label{s29}
\end{equation}
The system is always stable.

\section{Initial value problem for the linearized mean field Kramers equation}
\label{sec_k}

\subsection{The dielectric function}

We now consider the mean field Kramers equation
(\ref{g1}) which contains the Vlasov equation and the mean field Smoluchowski equation as particular cases. The Green function of the ordinary Kramers equation has
been computed by Chandrasekhar \cite{chandra}. It can be written as
\begin{equation}
G({\bf r}-{\bf r}_0,{\bf v},{\bf v}_0,t)=\frac{1}{(2\pi)^d}\frac{1}{(FG-H^2)^{d/2}}e^{-\frac{1}{2(FG-H^2)}(GR^2-2H{\bf R}\cdot {\bf S}+FS^2)},
\label{k1}
\end{equation}
with
\begin{equation}
F=\frac{D}{\xi^3}(2\xi t-3+4e^{-\xi t}-e^{-2\xi t}), \qquad G=\frac{D}{\xi}(1-e^{-2\xi t}),\qquad H=\frac{D}{\xi^2}(1-e^{-\xi t})^2,
\label{k2}
\end{equation}
\begin{equation}
{\bf R}={\bf r}-{\bf r}_0-\frac{1}{\xi}{\bf v}_0 (1-e^{-\xi t}),\qquad {\bf S}={\bf v}-{\bf v}_0e^{-\xi t}.
\label{k5}
\end{equation}
Its Fourier transform is
\begin{equation}
\hat{G}({\bf k},{\bf v},{\bf v}_0,t)=\frac{1}{(2\pi)^d}\frac{1}{(2\pi G)^{d/2}}e^{-\frac{1}{2G}(FG-H^2)k^2}e^{-\frac{i}{\xi}{\bf k}\cdot {\bf v}_0(1-e^{-\xi t})}e^{-\frac{S^2}{2G}}e^{-i\frac{H}{G}{\bf k}\cdot {\bf S}}.
\label{k7}
\end{equation}
We note that its Fourier-Laplace transform is not available in a
simple form contrary to the Fourier-Laplace transforms (\ref{v2}) and (\ref{s22}) of the Green
function of a free particle ($\xi=0$) and of an overdamped particle ($\xi\rightarrow +\infty$). Using Eq. (\ref{k7}), we obtain after
some calculations
\begin{equation}
\hat{Q}({\bf k},{\bf v}_0,t)=-\frac{1}{(2\pi)^d}e^{-\frac{1}{2}Fk^2}e^{-\frac{i}{\xi}{\bf k}\cdot {\bf v}_0(1-e^{-\xi t})},
\label{k8}
\end{equation}
and
\begin{equation}
\hat{P}({\bf k},t)=-\frac{1}{\xi}(2\pi)^d \hat{u}(k)\rho k^2(1-e^{-\xi t})e^{-\frac{Dk^2}{\xi^2}t}e^{\frac{Dk^2}{\xi^3}(1-e^{-\xi t})}.
\label{k9}
\end{equation}
For $\xi\rightarrow 0$ we recover Eq. (\ref{v4b}) and for
$\xi\rightarrow +\infty$ we recover Eq. (\ref{s22b}). The Laplace
transform of $\hat{P}({\bf k},t)$ is
\begin{equation}
\tilde{P}({\bf k},\omega)=\int_0^{+\infty}e^{i\omega t}\hat{P}({\bf k},t)\, dt.
\label{k10}
\end{equation}
Substituting Eq. (\ref{k9}) in Eq. (\ref{k10}) and making the change of variables $s=e^{-\xi t}$ for $\xi>0$ we can express the integral in terms of the incomplete Gamma functions
\begin{equation}
\gamma(\alpha,x)=\int_0^x e^{-t}t^{\alpha-1}\, dt.
\label{k11}
\end{equation}
We find
\begin{equation}
\tilde{P}({k},\omega)=-(2\pi)^d \hat{u}(k)\rho\beta m F\left (\frac{Dk^2}{\xi^3}-\frac{i\omega}{\xi},\frac{Dk^2}{\xi^3}\right ),
\label{k12}
\end{equation}
where we have defined
\begin{equation}
F(\alpha,x)=\frac{e^x}{x^{\alpha-1}}\left\lbrack\gamma(\alpha,x)-\frac{1}{x}\gamma(\alpha+1,x)\right\rbrack.
\label{k13}
\end{equation}
Some properties of this function are given in Appendix \ref{sec_f}.
The dielectric function can finally be written as
\begin{equation}
\epsilon({k},\omega)=1+(2\pi)^d\hat{u}(k)\rho\beta m F\left (\frac{Dk^2}{\xi^3}-\frac{i\omega}{\xi},\frac{Dk^2}{\xi^3}\right ).
\label{k15}
\end{equation}
In the strong friction limit $\xi\rightarrow +\infty$, 
using $F(\alpha,x)\sim x/[\alpha(\alpha+1)]$ for $x\rightarrow 0$ (see Appendix
\ref{sec_f}), we recover the dielectric function (\ref{s9}) associated with the
linearized Smoluchowski equation. In order to take the no friction limit, it
seems preferable to come back to the expression (\ref{k9}) of the polarization
function which reduces to Eq. (\ref{v4b}) when $\xi\rightarrow 0$. Substituting
this expression in Eq. (\ref{k10}) and using Eq. (\ref{g24}), we get Eq.
(\ref{alt1}) which is equivalent to the dielectric function (\ref{v4}-b)
associated with the linearized Vlasov equation (see Appendix \ref{sec_alt}).

\subsection{The dispersion relation and the stability criterion}
\label{sec_drv}

The dispersion relation $\epsilon({k},\omega)=0$ can be written as
\begin{equation}
1+(2\pi)^d\hat{u}(k)\rho\beta m F\left (\frac{Dk^2}{\xi^3}-\frac{i\omega}{\xi},\frac{Dk^2}{\xi^3}\right )=0.
\label{k17}
\end{equation}
The neutral mode corresponds to $\omega=0$. Using $F(x,x)=1$ which immediately results from Eq. (\ref{k14}), we get the condition $1+(2\pi)^d\hat{u}(k)\rho\beta m=0$. It can be shown that the system is stable with respect to a perturbation with wavenumber $k$ when 
\begin{equation}
1+(2\pi)^d\hat{u}(k)\rho\beta m>0,
\label{k19}
\end{equation}
and unstable otherwise. This stability criterion can also be obtained from
the study of the second order variations of the free energy (see Appendix
\ref{sec_so}). We stress 
that the stability criterion (\ref{k19}) does {\it not} depend on the friction
coefficient $\xi$ while, of course, the evolution of the perturbation in the
stable and unstable regimes (i.e. the value of the complex pulsations that are
the solution of the dispersion relation) depend on it in a non trivial manner.

{\it Remark:} The non-interacting limit corresponds to $\hat{u}({k})=0$. In
that case, it is necessary that $F(.)\rightarrow +\infty$ in Eq. (\ref{k17}).
According to Eq. (\ref{k14}), this implies that $\omega=i\omega_i$ with
$\omega_i=-Dk^2/\xi^2-n\xi$ where $n\ge 0$ is any positive integer. We therefore
recover the well-known proper pulsations of the usual Kramers equation
\cite{risken}.

\subsection{Application to the BMF model, self-gravitating systems, and plasmas}
\label{sec_cor}

\begin{figure*}
\centering
\includegraphics[width=0.5\textwidth]{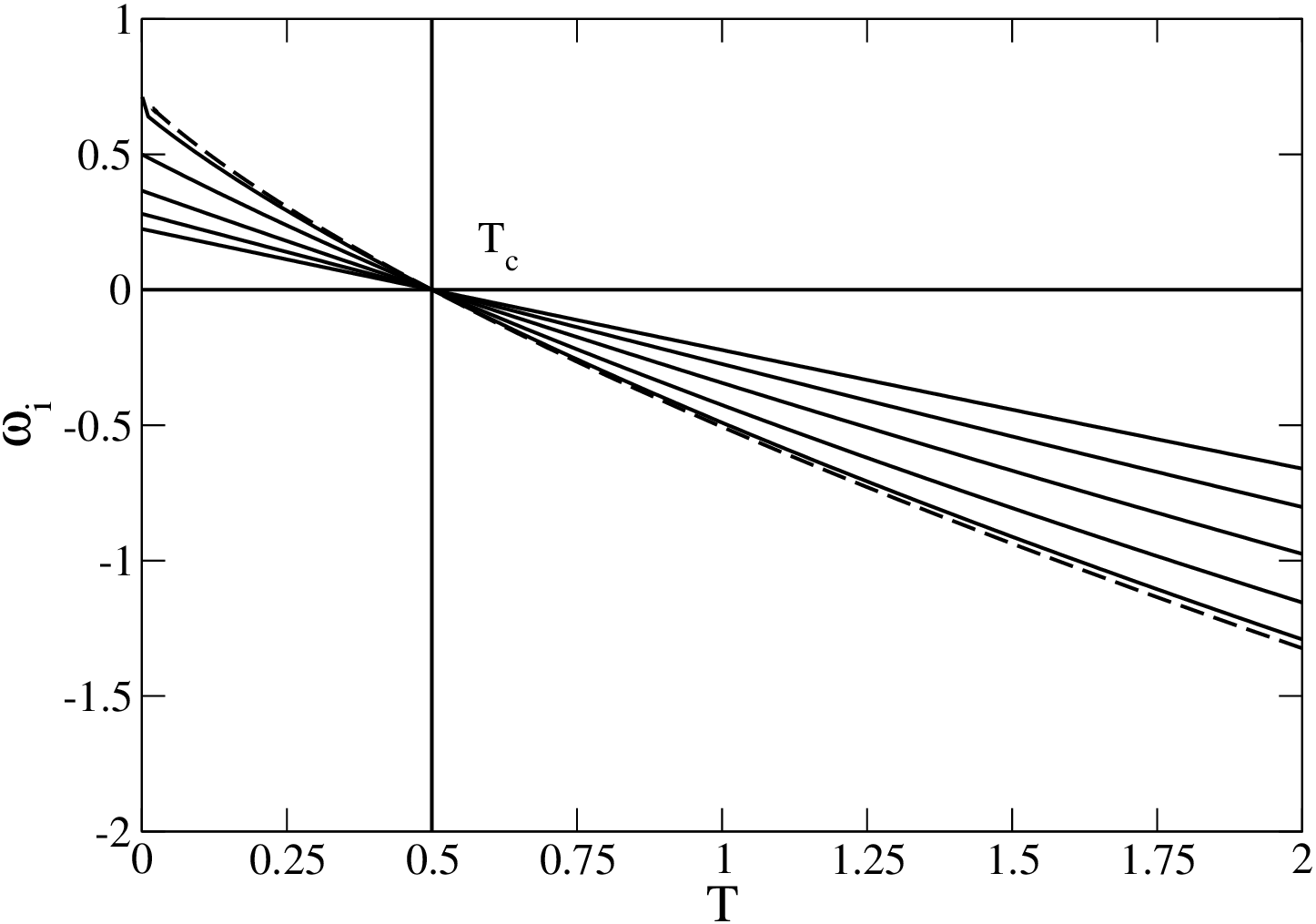}
\caption{Complex pulsation $\omega=i\omega_i$ as a function of the 
temperature $T$ for different values of the friction parameter $\xi=0.1, 0.5, 1,
1.5, 2$ in the case of the attractive BMF model described by the mean field
Kramers equation. We have plotted only the fundamental pulsation. The dashed
line, obtained from Eq. (\ref{v30}), corresponds to the dissipationless limit
$\xi=0$ (Vlasov). For $\xi\rightarrow +\infty$ (Smoluchowski) the complex
pulsation is given by Eq. (\ref{s27}-b).}
\label{wit}
\end{figure*}

For the attractive BMF model, using
$\hat{u}_n=\frac{1}{2N}(2\delta_{n,0}-\delta_{n,1}-\delta_{n,-1})$ and
$\rho=1/(2\pi)$, and considering the modes $n=\pm 1$ (the modes $n\neq \pm 1$ evolve with the proper pulsations of the usual Kramers equation), the dispersion relation (\ref{k17}) can be written as
\begin{equation}
1-\frac{1}{2T} F\left (\frac{T}{\xi^2}-\frac{i\omega}{\xi},\frac{T}{\xi^2}\right )=0.
\label{k23}
\end{equation}
The system is stable if $T>T_c=1/2$ and unstable (with respect to the modes $n=\pm 1$) if $T<T_c$. For $\xi\gg 1$, using Eq. (\ref{k14b}), we get 
\begin{equation}
i\xi\omega\simeq \left (1-\frac{1}{2\xi^2}\right )\left (T-\frac{1}{2}\right ).
\label{k23c}
\end{equation}
This is the first order correction to the Smoluchowski limit $\xi\rightarrow +\infty$. We now assume  that $\omega=i\omega_i$ (for $T<T_c$, this is the only solution of the dispersion relation with $\omega_i>0$). For $T\rightarrow 0$, using Eq. (\ref{k14b}), we obtain $\omega_i(T,\xi)\simeq \omega_i(0,\xi)+b(\xi)T+...$ with
\begin{equation}
\omega_i(0,\xi)=\frac{-\xi\pm\sqrt{2+\xi^2}}{2},\qquad b(\xi)=\frac{1}{\xi}\left (\frac{2}{\pm 3\xi\sqrt{2+\xi^2}+2+\xi^2}-1\right ).
\label{k23a}
\end{equation}
Close to the neutral mode $\omega_i=0$, i.e. for $T\rightarrow T_c=1/2$, using Eq. (\ref{f1}) we obtain
\begin{equation}
\omega_i(T,\xi)=\frac{\xi}{G\left (\frac{1}{2\xi^2}\right )}(1-2T).
\label{k23b}
\end{equation}
For $\xi\rightarrow 0$ and $\xi\rightarrow +\infty$, Eqs. (\ref{k23a}) and (\ref{k23b}) return the results of Sects. \ref{sec_v} and \ref{sec_s} respectively. The fundamental pulsation $\omega_i$ is plotted as a function of the temperature $T$ in Fig. \ref{wit} for different values of the friction parameter $\xi$. For $T<T_c$, $\omega_i>0$ so the perturbation grow exponentially rapidly without oscillating. For $T>T_c$, $\omega_i<0$ so the perturbation decreases exponentially rapidly without oscillating (in that case, other modes exist with a non-vanishing pulsation $\omega_r$ but they are damped more rapidly).

\begin{figure*}
\centering
\includegraphics[width=0.5\textwidth]{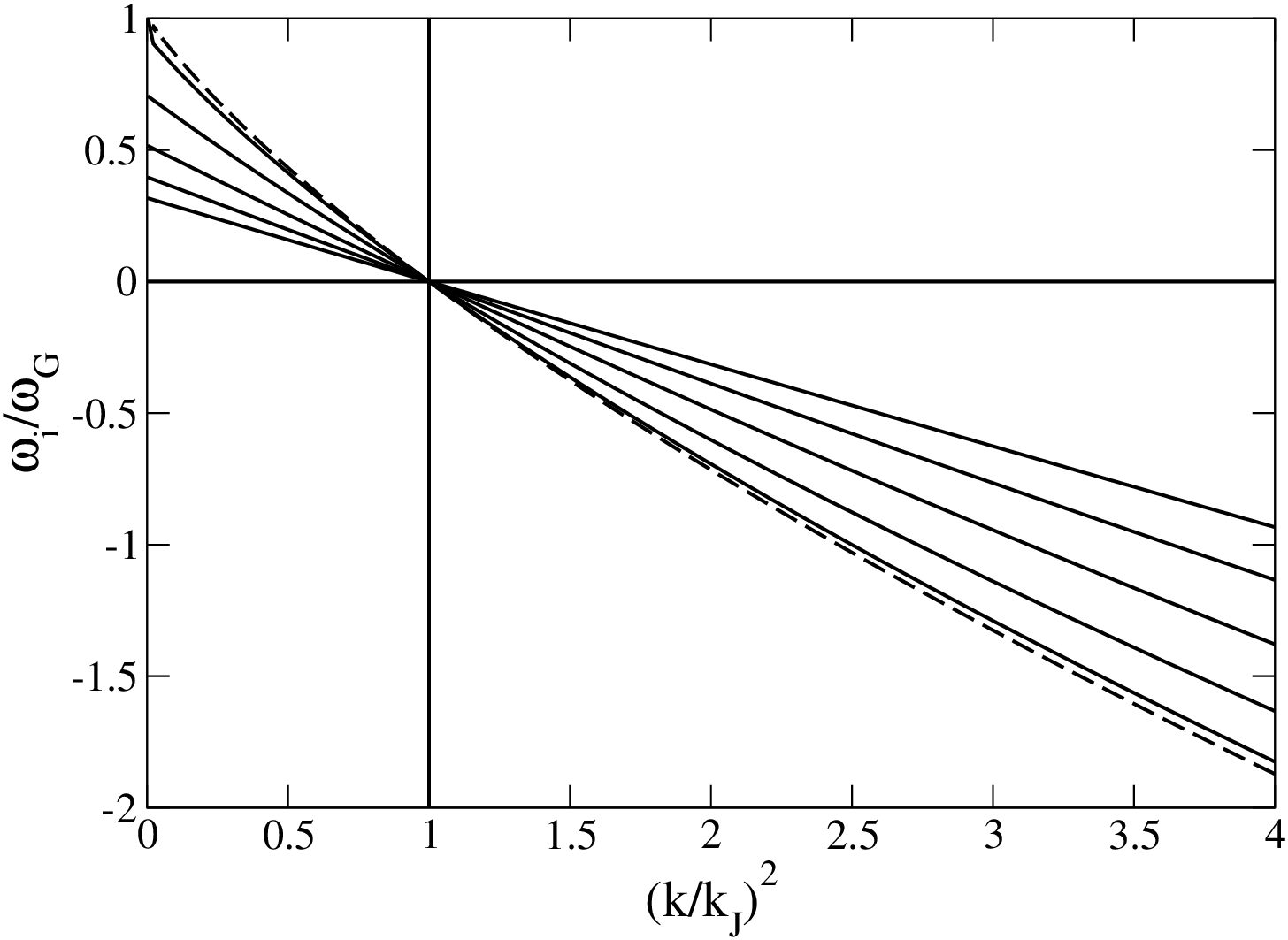}
\caption{Complex pulsation $\omega=i\omega_i$ as a function of the wavenumber $k/k_J$ for different values of  $1/(\sqrt{2}{\cal N})=0.1, 0.5, 1, 1.5, 2$ in the case of self-gravitating Brownian systems described by the mean field Kramers equation. We have plotted only the fundamental pulsation. The dashed line, obtained from Eq. (\ref{v25}), corresponds to the dissipationless limit $\xi=0$ (Vlasov). For $\xi\rightarrow +\infty$ (Smoluchowski) the complex pulsation is given by Eq. (\ref{s26}).}
\label{witGRAVITE}
\end{figure*}

For self-gravitating systems, using $(2\pi)^d \hat{u}(k)=-S_d G/k^2$, the dispersion relation (\ref{k17}) can be written as
\begin{equation}
1-\frac{k_J^2}{k^2} F\left ({\cal N}^2 \frac{k^2}{k_J^2}-{\cal N}\frac{i\omega}{\omega_G},{\cal N}^2 \frac{k^2}{k_J^2}\right )=0.
\label{k22}
\end{equation}
We have introduced the dimensionless number ${\cal N}=t_B/t_D=\omega_G/\xi$ corresponding to the ratio between the Brownian time and the dynamical time. The system is stable if $k>k_J$ and unstable if $k<k_J$. For ${\cal N}\ll 1$, using Eq. (\ref{k14b}), we get
\begin{equation}
i\xi\omega\simeq (1-{\cal N}^2)(c_s^2k^2-S_dG\rho).
\label{k23cb}
\end{equation}
This is the first order correction to the Smoluchowski limit ${\cal N}\rightarrow 0$. We now assume  that $\omega=i\omega_i$ (for $k<k_J$, this is the only solution of the dispersion relation with $\omega_i>0$) For $k\ll k_J$, using Eq. (\ref{k14b}), we obtain $\omega_i(k/k_J,{\cal N})/\omega_G\simeq \omega_i(0,{\cal N})/\omega_G+B({\cal N})(k/k_J)^2+...$ with
\begin{equation}
\frac{\omega_i(0,{\cal N})}{\omega_G}=\frac{-\frac{1}{\cal N}\pm\sqrt{4+\frac{1}{{\cal N}^2}}}{2},\qquad B({\cal N})={\cal N}\left (\frac{2}{\pm\frac{3}{\cal N}\sqrt{1+\frac{1}{4{\cal N}^2}}+2+\frac{1}{2{\cal N}^2}}-1\right ).
\label{k23ab}
\end{equation}
Close to the neutral mode $\omega_i=0$, i.e. for $k\rightarrow k_J$, using Eq. (\ref{f1}) we obtain
\begin{equation}
\frac{\omega_i(k/k_J,{\cal N})}{\omega_G}=\frac{1}{{\cal N}G({\cal N}^2)}\left (1-\frac{k^2}{k_J^2}\right ).
\label{k23bb}
\end{equation}
For ${\cal N}\rightarrow +\infty$ and ${\cal N}\rightarrow 0$, Eqs. (\ref{k23ab}) and (\ref{k23bb}) return the results of Sects. \ref{sec_v} and \ref{sec_s} respectively.  The fundamental pulsation $\omega_i$ is plotted as a function of the wavenumber $k/k_J$ in Fig. \ref{witGRAVITE} for different values of ${\cal N}$. For $k<k_J$, $\omega_i>0$ so the perturbation grow exponentially rapidly without oscillating. For $k>k_J$, $\omega_i<0$ so the perturbation decreases exponentially rapidly without oscillating (in that case, other modes exist with a non-vanishing pulsation $\omega_r$ but they are damped more rapidly). Actually, the description of self-gravitating systems is similar to the description of the attractive BMF model provided that we make the correspondences $\omega\leftrightarrow \omega/(\sqrt{2}\omega_G)$, $T\leftrightarrow k^2/(2k_J^2)$, and $\xi\leftrightarrow 1/(\sqrt{2}{\cal N})$.

\begin{figure*}
\centering
\includegraphics[width=0.5\textwidth]{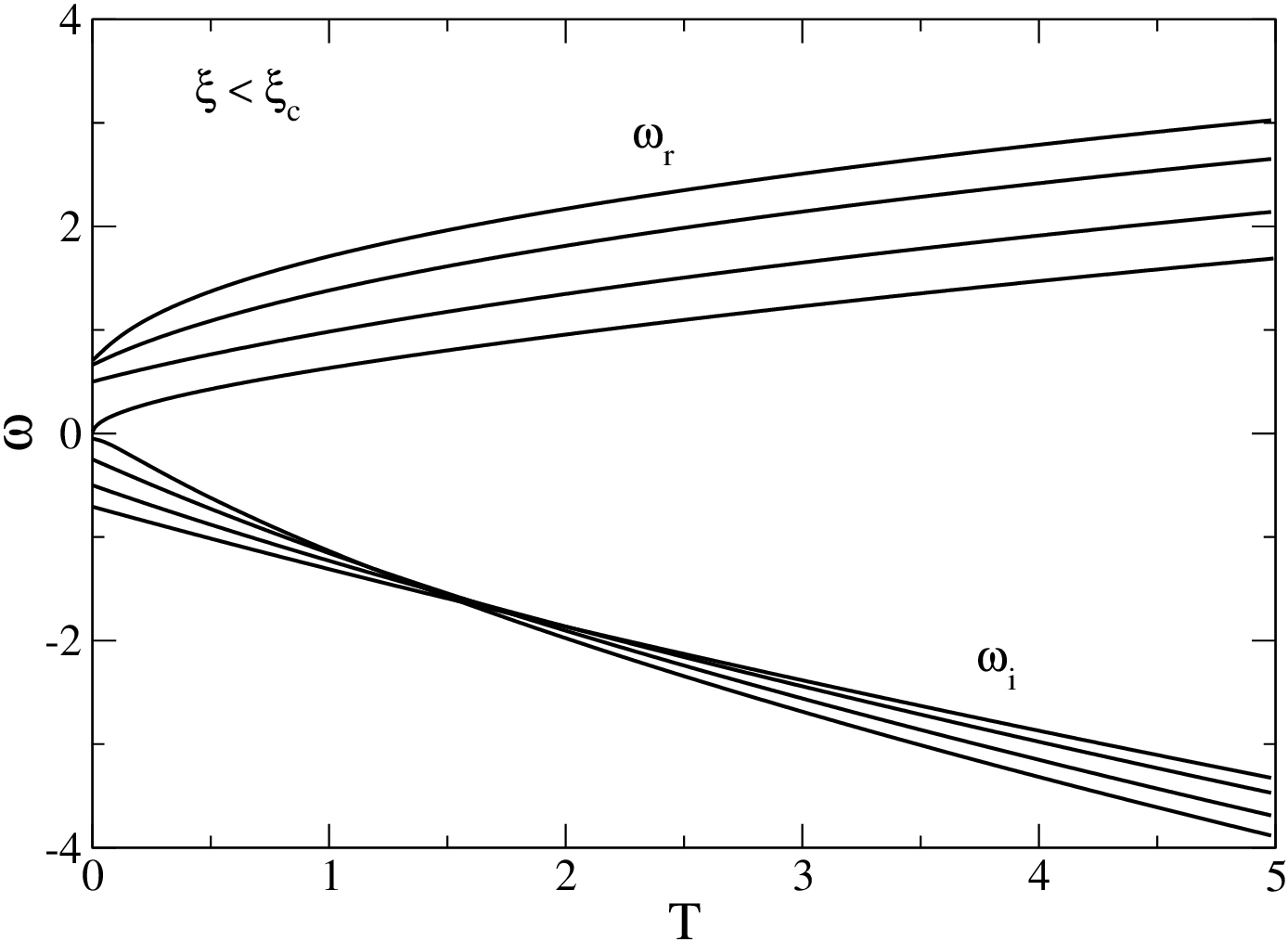}
\caption{Real and imaginary parts of the complex pulsation $\omega$ as a function of the temperature $T$ for different values of the friction parameter $\xi=0.1, 0.5, 1, \sqrt{2}$ ($\xi\le \xi_c$) in the case of the repulsive BMF model described by the mean field Kramers equation. The ordering of the curves may be seen by considering the case $T=0$. For $\xi=0$ (Vlasov),  $\omega_r=1/\sqrt{2}$ and $\omega_i=0$. For $\xi=\xi_c$ $\omega_r=0$ and $\omega_i=-1/\sqrt{2}$.}
\label{witREPULSIVE}
\end{figure*}

\begin{figure*}
\centering
\includegraphics[width=0.5\textwidth]{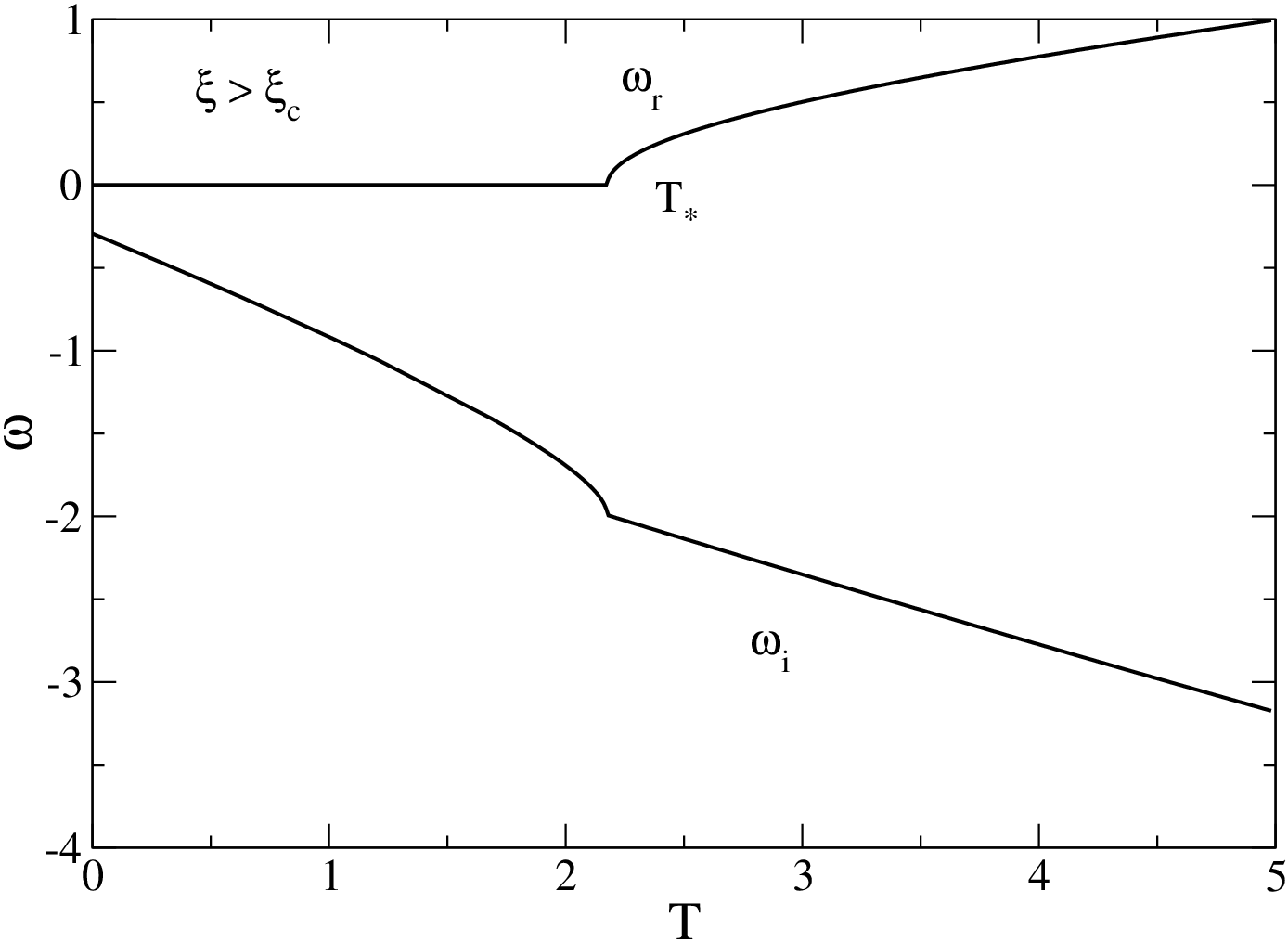}
\caption{Real and imaginary parts of the complex pulsation $\omega$ as a function of the temperature $T$ for $\xi=2$ ($\xi>\xi_c$) in the case of the repulsive BMF model described by the mean field Kramers equation. We find that  $\omega_r=0$ for $T<T_c(\xi)$ and $\omega_r\neq 0$ for $T>T_c(\xi)$. This is similar to a second order phase transition. For $\xi\rightarrow +\infty$ (Smoluchowski), the critical temperature $T_c(\xi)$ is  rejected to infinity and the complex pulsation is given by Eq. (\ref{s30}-b).   }
\label{Firstorder}
\end{figure*}

For the repulsive BMF model, using
$\hat{u}_n=-\frac{1}{2N}(2\delta_{n,0}-\delta_{n,1}-\delta_{n,-1})$
and $\rho=1/(2\pi)$, and considering the modes $n=\pm 1$ (the modes $n\neq \pm 1$ evolve with the proper pulsations of the usual Kramers equation), the dispersion relation (\ref{k17}) can be written as
\begin{equation}
1+\frac{1}{2T} F\left (\frac{T}{\xi^2}-\frac{i\omega}{\xi},\frac{T}{\xi^2}\right )=0.
\label{k25}
\end{equation}
The system is always stable.  For $\xi\gg 1$, using Eq. (\ref{k14b}), we get
\begin{equation}
i\xi\omega\simeq \left (1+\frac{1}{2\xi^2}\right )\left (T+\frac{1}{2}\right ).
\label{k23cf}
\end{equation}
This is the first order correction to the Smoluchowski limit $\xi\rightarrow +\infty$. For $T\rightarrow 0$, using Eq. (\ref{k14b}), we obtain $\omega(T,\xi)\simeq \omega(0,\xi)+b(\xi)T+...$ with
\begin{equation}
\omega(0,\xi)=\frac{-i\xi\pm\sqrt{2-\xi^2}}{2},\qquad b(\xi)=\frac{i}{\xi}\left (\frac{2}{\pm 3i\xi\sqrt{2-\xi^2}+2-\xi^2}-1\right ).
\label{grle}
\end{equation}
Let us consider the case $T=0$. Eq. (\ref{grle}-a) shows that there is
a critical friction parameter $\xi_c=\sqrt{2}$. For $\xi<\sqrt{2}$,
the perturbation oscillates with a pulsation
$\omega_r=\frac{1}{2}\sqrt{2-\xi^2}$ and is damped at a rate
$\omega_i=-\xi/2$. For $\xi>\sqrt{2}$, the perturbation is damped at a
rate $\omega_i=-\xi/2+\frac{1}{2}\sqrt{\xi^2-2}$ without oscillating.  For
$\xi\rightarrow
+\infty$, we recover the results of Sect. \ref{sec_s}. For
$\xi\rightarrow 0$, we recover the results of Sect. \ref{sec_v} for
the real part of the complex pulsation (\ref{v38}-a), but we do not
obtain the Landau damping (\ref{v38}-b).
Therefore, we conclude that frictional effects erase the Landau
damping.\footnote{We recall that our expansion is valid for fixed
$\xi>0$ and $T\rightarrow 0$.  On the other hand, for fixed $T$ and
$\xi\rightarrow 0$, we recover the results of Landau \cite{landau} since the
dispersion relation coincides with the one obtained from the
linearized Vlasov equation (see the comment after
Eq. (\ref{k9})). Therefore, the limits $\xi\rightarrow 0$ and
$T\rightarrow 0$ do not commute (see the Appendix \ref{sec_xizero}
where we consider $\xi\rightarrow 0$ then $T\rightarrow 0$ while here
we have considered $T\rightarrow 0$ then $\xi\rightarrow
0$). Indeed, there is an indetermination when both $T$
and $\xi$ go to zero since the ratio $T/\xi$ is not well-defined.} The real and
imaginary parts of the fundamental pulsation
$\omega$ are plotted as a function of the temperature $T$ in
Figs. \ref{witREPULSIVE} and
\ref{Firstorder} for different values of the friction parameter
$\xi$. For $\xi<\xi_c$, the perturbation oscillates with a pulsation
$\omega_r\neq 0$ and is damped at a rate $\omega_i<0$. For
$\xi>\xi_c$, there exist a critical temperature $T_*(\xi)$ such that
$\omega_r=0$ for $T<T_*(\xi)$ and $\omega_r\neq 0$ for
$T>T_*(\xi)$. This is similar to a second order phase transition. The
derivatives of $\omega_r(T,\xi)$ and $\omega_i(T,\xi)$ presents a
discontinuity at $T=T_*(\xi)$.

For Coulombian plasmas, using $(2\pi)^d \hat{u}(k)=S_d e^2/m^2k^2$,
the dispersion relation (\ref{k17}) can be written as
\begin{equation}
1+\frac{k_D^2}{k^2} F\left ({\cal N}^2 \frac{k^2}{k_D^2}-{\cal N}\frac{i\omega}{\omega_P},{\cal N}^2 \frac{k^2}{k_D^2}\right )=0.
\label{k24}
\end{equation}
We have introduced the dimensionless number ${\cal N}=t_B/t_D=\omega_P/\xi$
corresponding to the ratio between the Brownian time and the dynamical time. The
system is always stable. For ${\cal N}\ll 1$, using Eq. (\ref{k14b}), we get
\begin{equation}
i\xi\omega\simeq (1+{\cal N}^2)(c_s^2k^2+\omega_P^2).
\label{k23cbz}
\end{equation}
This is the first order correction to the Smoluchowski limit ${\cal
N}\rightarrow 0$. For $k\ll k_D$, using Eq. (\ref{k14b}), we obtain
$\omega(k/k_D,{\cal N})/\omega_P\simeq \omega(0,{\cal N})/\omega_P+B({\cal
N})(k^2/k_D^2)+...$ with
\begin{equation}
\frac{\omega(0,\xi)}{\omega_P}=\frac{-\frac{i}{\cal N}\pm\sqrt{4-
\frac{1}{{\cal N}^2}}}{2},\qquad B({\cal N})=i{\cal N}\left (\frac{2}{\pm\frac{3}{\cal N}i\sqrt{1-\frac{1}{4{\cal N}^2}}+2-\frac{1}{2{\cal N}^2}}-1\right ).
\label{grlep}
\end{equation}
Let us consider the case $k=0$.  Eq. (\ref{grlep}-a) shows that there is a
critical number ${\cal N}_c=1/2$. For ${\cal N}>1/2$, the perturbation
oscillates with a pulsation $\omega_r/\omega_P=\frac{1}{2}\sqrt{4-1/{\cal N}^2}$
and is damped at a rate $\omega_i/\omega_P=-1/(2{\cal N})$. For ${\cal N}<1/2$,
the perturbation  is damped at a rate $\omega_i/\omega_P=-1/(2{\cal
N})+\frac{1}{2}\sqrt{1/{\cal N}^2-4}$ without oscillating. For ${\cal
N}\rightarrow 0$, we recover the results of Sect. \ref{sec_s}. For ${\cal
N}\rightarrow +\infty$, we recover the results of Sect. \ref{sec_v} for the real
part of the complex pulsation (\ref{v35}-a), but we do not obtain the Landau
damping (\ref{v38}-b).  Therefore, we
conclude that frictional effects erase the Landau damping (see footnote 4).
The description of plasmas is similar to the description of the repulsive BMF
model provided that we make the correspondences $\omega\leftrightarrow
\omega/(\sqrt{2}\omega_P)$, $T\leftrightarrow k^2/(2k_D^2)$, and
$\xi\leftrightarrow 1/(\sqrt{2}{\cal N})$. Therefore, the evolution of the real
and imaginary parts of the complex pulsation as a function of the wavenumber
$k/k_D$ for different values of the friction can be easily deduced from Figs. 
\ref{witREPULSIVE} and \ref{Firstorder}. The same phenomenon of ``first order
phase transition'' occurs at a particular wavenumber $k_*(\xi)$ when  ${\cal
N}<1/2$.

\subsection{Graphical construction to locate the purely imaginary pulsations}
\label{sec_graph}

\begin{figure*}
\centering
\includegraphics[width=0.5\textwidth]{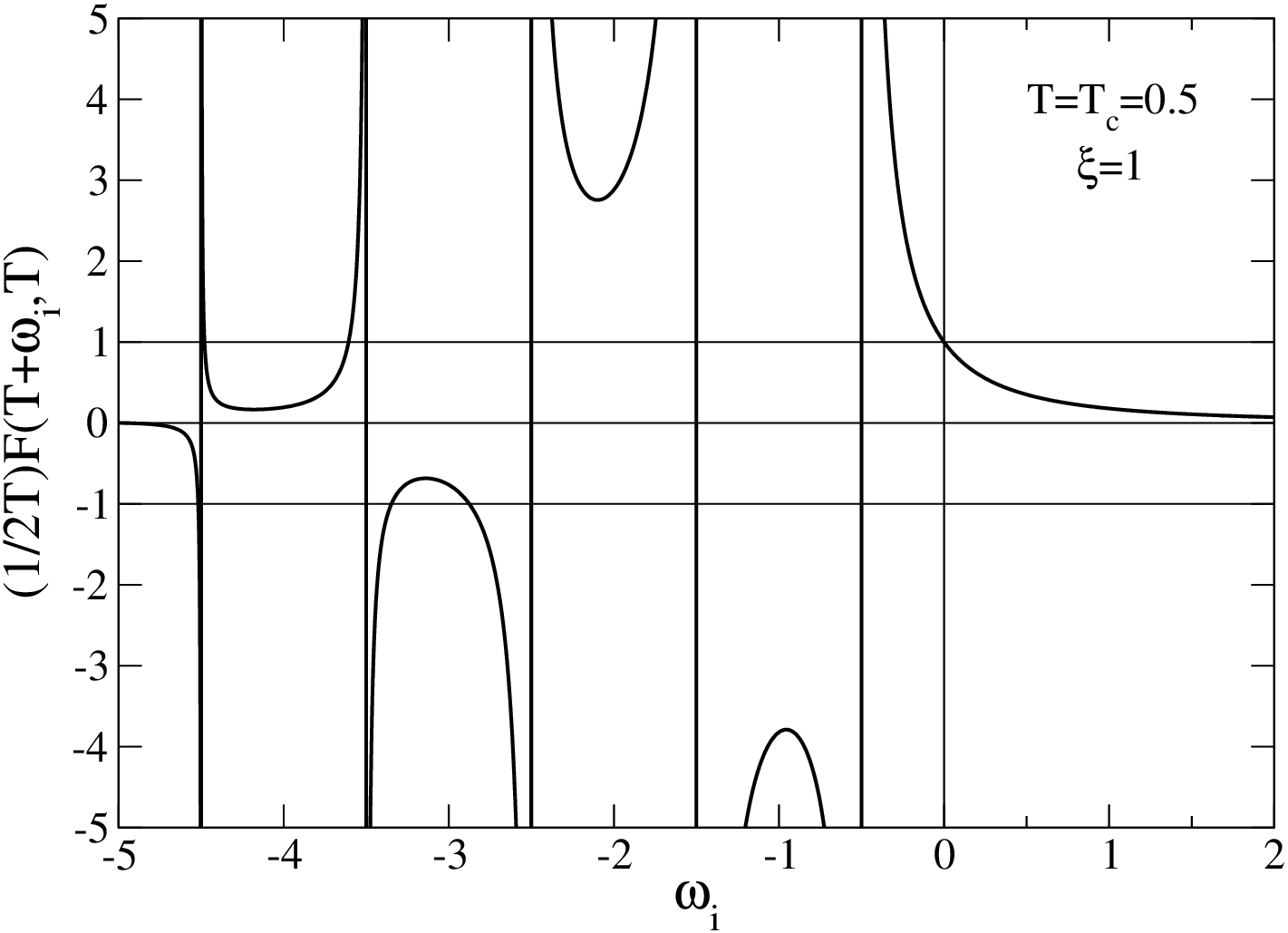}
\caption{Graphical construction determining the purely imaginary pulsations $\omega=i\omega_i$ of the BMF model for $T=T_c=1/2$ and $\xi=1$. They are given by the intersections between the curve and the horizontal line $+1$ (attractive) or $-1$ (repulsive). In the attractive case, the fundamental pulsation is $\omega=0$ (neutral).}
\label{F1t0p5}
\end{figure*}

\begin{figure*}
\centering
\includegraphics[width=0.5\textwidth]{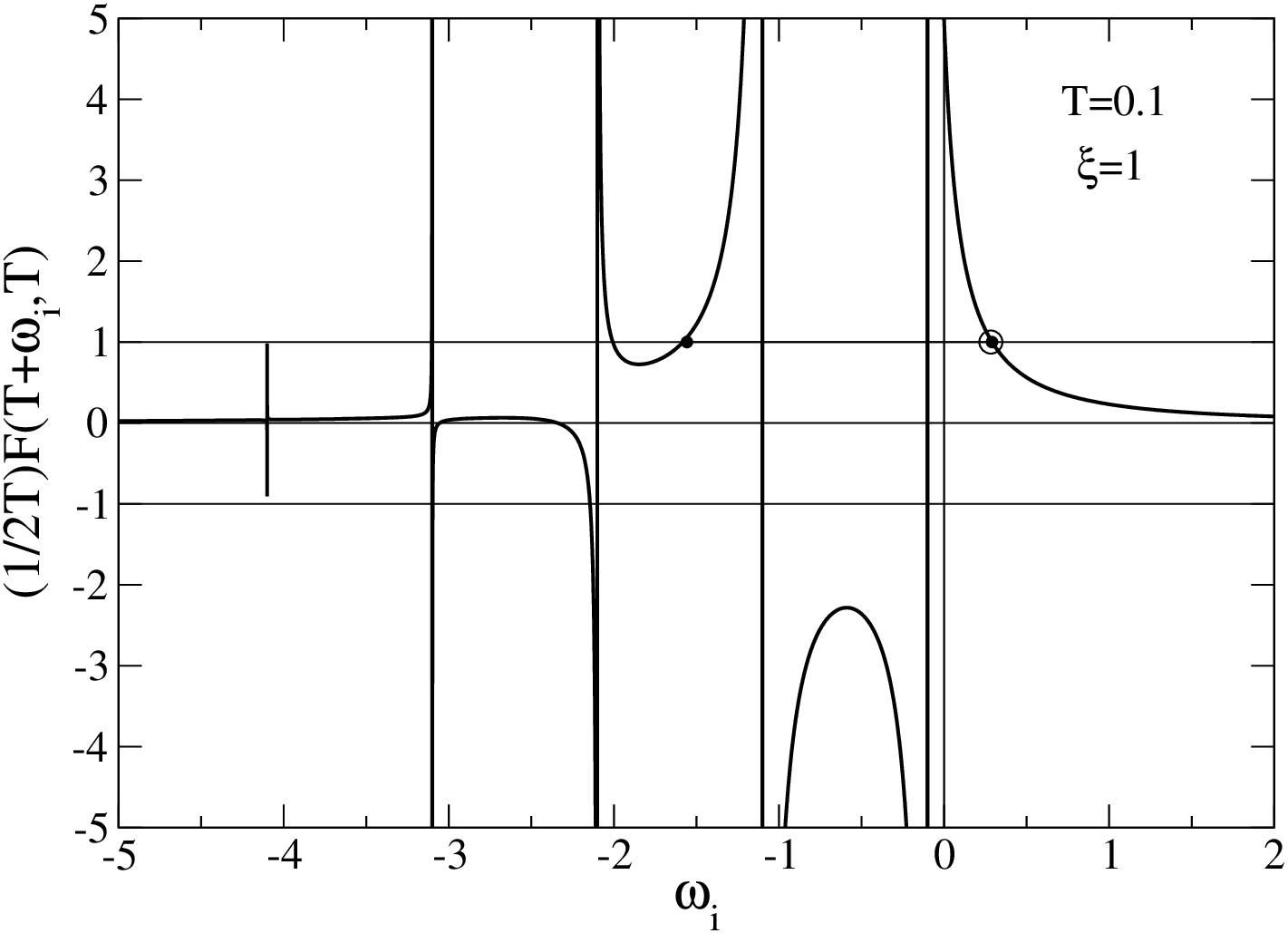}
\caption{Graphical construction determining the purely imaginary pulsations $\omega=i\omega_i$ of the BMF model for $T=0.1<T_c$ and $\xi=1$. In the attractive case, the imaginary part of the fundamental pulsation is $\omega_i>0$ (unstable). The black bullets correspond to the analytical values (\ref{k23a}) valid for $T\rightarrow 0$ and the white bullet to the analytical value (\ref{k23b}) valid for $T\rightarrow T_c$. In the repulsive case, there is no purely imaginary pulsation since $\xi<\xi_c=\sqrt{2}$ (see the text for more details).}
\label{F1t0p1}
\end{figure*}

\begin{figure*}
\centering
\includegraphics[width=0.5\textwidth]{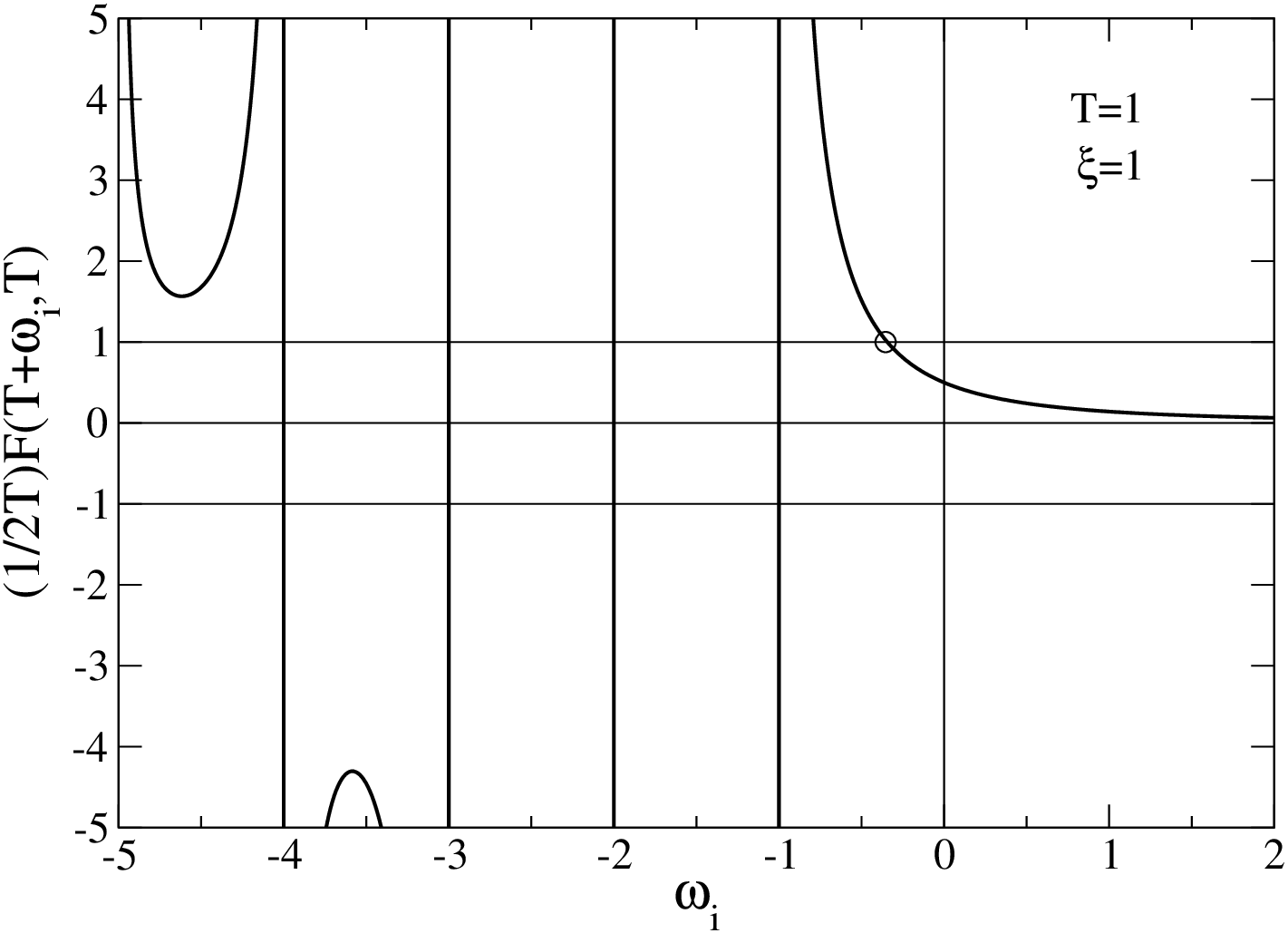}
\caption{Graphical construction determining the purely imaginary pulsations $\omega=i\omega_i$ of the BMF model for $T=1>T_c$ and $\xi=1$. In the attractive case, the imaginary part of the fundamental pulsation is $\omega_i<0$ (stable). The white bullet corresponds to the analytical value (\ref{k23b}) valid for $T\rightarrow T_c$.}
\label{F1t1}
\end{figure*}

For the attractive and repulsive BMF models\footnote{We treat here the case of the BMF model but the discussion is similar for self-gravitating systems and plasmas provided that we use the correspondences given in Sec. \ref{sec_cor}.}, the dispersion relation may be written as
\begin{equation}
\frac{1}{2T}F(T+\omega_i,T)=\pm 1.
\label{disp1}
\end{equation}
To simplify the discussion we have taken $\xi=1$ but we shall explain later how to treat the general case. We have also assumed that $\omega=i\omega_i$. To understand the structure of the dispersion relation, we have plotted $\frac{1}{2T}F(T+\omega_i,T)$ as a function of $\omega_i$ for different values of the temperature in Figs. \ref{F1t0p5}-\ref{F1t1}. As explained previously, the function $F(T+\omega_i,T)$ diverges when $\omega_i=-T-n$ for any integer $n\ge 0$. The zeros $\omega_i$ correspond to the intersections between the curve $\frac{1}{2T}F(T+\omega_i,T)$ and the horizontal line $+1$ in the attractive case or the horizontal line $-1$ in the repulsive case.  Depending on the value of the temperature, there may be several intersections corresponding to purely imaginary pulsations $\omega=i\omega_i$. On the other hand, the absence of intersection may reveal that the pulsation has a real part: $\omega=\omega_r+i\omega_i$ with $\omega_r\neq 0$. Of course, the pulsation with the highest imaginary part (fundamental pulsation) is the most relevant.

Let us first consider the attractive case. Since the curve
$\frac{1}{2T}F(T+\omega_i,T)$ tends to $+\infty$ when $\omega_i\rightarrow -T$
and to zero when $\omega_i\rightarrow +\infty$ (see Appendix \ref{sec_f}), we
conclude that there is always a solution $\omega_i=i\omega_i$ with $\omega_i\ge
-T$. This is the fundamental pulsation. For $T=T_c=1/2$ we have  $\omega_i=0$
(neutral), for $T<T_c$ we have $\omega_i>0$ (unstable), and for $T>T_c$ we have
$\omega_i<0$ (stable). This is illustrated in Figs. \ref{F1t0p5}-\ref{F1t1}. The
evolution of the fundamental pulsation $\omega_i$ with $T$ is represented in
Fig. \ref{wit} for different values of $\xi$. In Fig. \ref{F1t0p1}, since the
temperature $T=0.1$ is small, we have represented by black bullets the values of
the pulsation given by the approximate expression (\ref{k23a}). We see that they
give a good agreement with the numerical (exact) values of the first two
pulsations with the highest imaginary part. We note that the other pulsations
are very close to the asymptotes at $\omega_i=-T-n$ with $n\ge 2$ where $F$
diverges. This explains why we cannot obtain them with the expansion that we
have used to obtain Eq. (\ref{k23a}). On the other hand,  since the temperatures
$T=0.1$ and $T=1$ in Figs. \ref{F1t0p1}  and \ref{F1t1} are not too far from
$T_c=1/2$ we have represented by a white bullet the value of the pulsation given
by the approximate expression (\ref{k23b}). Again, we obtain a good agreement
with the numerical (exact) value of the fundamental pulsation. We note that if
the temperature is sufficiently small, there exist other purely imaginary
pulsations. There also exist pulsations with a non-vanishing real part producing
damped oscillations. However, 
these pulsations are less ``fundamental'' than the pulsation represented in
Fig. \ref{wit} since they decay more rapidly.

We now consider the repulsive case. Since the function
$\frac{1}{2T}F(T+\omega_i,T)$ is positive for $\omega_i>-T$, there is
no intersection with the horizontal line $-1$ in that range. We
conclude therefore that $\omega_i$ is necessarily negative so that the
system is always stable. The fundamental pulsation has its imaginary
part $\omega_i$ in the range $-T-1\le \omega_i\le -T$. For $\xi=1$,
there is no intersection with the horizontal line $-1$ in that
range. This implies that the fundamental pulsation has a non-zero real
part $\omega_r\neq 0$. This is in agreement with the result
(\ref{grle}) valid for $T\rightarrow 0$ which shows that the pulsation
has a non-vanishing real part when $\xi<\xi_c=\sqrt{2}$. We note that,
depending on the temperature, there may exist purely imaginary
pulsations with $\omega_i\le -T-2$. For $T\rightarrow 0$, they are
very close to the asymptotes at $\omega_i=-T-n$ with $n\ge 2$ which
explains why we cannot obtain them with the expansion that we have
used to obtain Eq. (\ref{grle}). For $\xi>\xi_c=\sqrt{2}$, the curve
$\frac{1}{2T}F(T+\omega_i,T)$ intersects the horizontal line $-1$ in
the range $-T-1\le \omega_i\le -T$ provided that the temperature is
not too high. In that case, the fundamental pulsation is purely
imaginary. For $T\rightarrow 0$ it is given by Eq. (\ref{grle}). These
results can be understood graphically as follows. First, restoring the
friction parameter, we note that the dispersion relation may be
written as
$\frac{\xi^2}{2T}F(T/\xi^2+\omega_i/\xi,T/\xi^2)=-\xi^2$. Therefore,
the curves of Fig. \ref{F1t0p5}-\ref{F1t1} correspond to the left hand
side of this relation provided that $T$ is interpreted as $T/\xi^2$
and $\omega_i$ is interpreted as $\omega_i/\xi$. In that case, we have
to consider the intersection with these curves and the horizontal line
$-\xi^2$. For fixed $\xi$ and $T\rightarrow 0$, the maximum of the
curve in the range $-1\le \omega_i\le 0$ is $-2$ (the curve is
$1/[2(\omega_i/\xi)(\omega_i/\xi+1)]$). This implies that for $T=0$ the
fundamental pulsation is purely imaginary when $\xi>\sqrt{2}$
(intersection) while it has a non-vanishing real part when
$\xi<\sqrt{2}$ (no intersection). Furthermore, for fixed $\xi$, we see
that the maximum of the curve
$\frac{\xi^2}{2T}F(T/\xi^2+\omega_i/\xi,T\xi^2)$ in the range $-T-1\le
\omega_i\le -T$ decreases as $T$ increases. Therefore, when
$\xi>\sqrt{2}$ the fundamental pulsation is purely imaginary for
$T<T_c(\xi)$ (intersection) while it has a non-vanishing real part
when $T>T_c(\xi)$ (no intersection) in agreement with the discussion
of Sec. \ref{sec_cor}.

\section{The mean field damped Euler equations}
\label{sec_e}

It is interesting to compare the results obtained from the mean field Kramers
equation with those obtained from 
the mean field damped Euler equations which also include a dissipative term
\cite{gen,nfp,lr}. However, we stress that the damped Euler equations, which
rely on a local thermodynamic equilibrium (LTE) assumption, cannot be rigorously
derived from the Kramers equation \cite{hydrobrown}. Therefore, the dispersion
relation associated with  the linearized mean field Kramers equation is very
different from the dispersion relation associated with  the mean field damped
Euler equations except in particular limits.

\subsection{The local thermodynamic equilibrium assumption}

The mean field damped Euler equations write
\begin{eqnarray}
{\partial\rho\over\partial
t}+\nabla\cdot (\rho {\bf u})=0, \qquad\frac{\partial {\bf u}}{\partial t}+({\bf
u}\cdot \nabla){\bf u}=-\frac{1}{\rho}\nabla p-\nabla\Phi-\xi{\bf u},\label{e1}
\end{eqnarray}
\begin{equation}
\Phi({\bf r},t)=\int u({\bf r}-{\bf r}')\rho({\bf r}',t)\, d{\bf r}'.
\label{e3}
\end{equation}
For $\xi=0$, we recover the mean field Euler equations and for $\xi\rightarrow +\infty$ we recover the mean field Smoluchowski equation (\ref{smf10gbis}). The mean field damped Euler equations with an isothermal equation of state (\ref{piso}) may be obtained by taking the hydrodynamic moments of the mean field Kramers equation and making a LTE assumption
\begin{equation}
f({\bf r},{\bf v},t)=\left (\frac{\beta m}{2\pi}\right )^{d/2}\rho({\bf r},t)
e^{-\frac{1}{2}\beta m [{\bf v}-{\bf u}({\bf r},t)]^2},
\label{e4}
\end{equation}
to close the hierarchy of equations. A generalization of this procedure to treat systems described by other equations of state is developed in \cite{gen,nfp,longshort}. However, we stress that there is no rigorous justification of the LTE assumption \cite{hydrobrown} except in the strong friction limit $\xi\rightarrow +\infty$ where we obtain the (generalized) mean field Smoluchowski equation.

In the isothermal case, the mean field damped Euler equation satisfies an $H$-theorem for the free energy
\begin{eqnarray}
\label{e5}
F[\rho,{\bf u}]=\int \rho \frac{{\bf u}^2}{2}\, d{\bf r}+{1\over 2}\int\rho\Phi \, d{\bf r}+k_B T\int \frac{\rho}{m}\ln\left (\frac{\rho}{Nm}\right )\, d{\bf r}-\frac{d}{2}N k_B T\ln \left (\frac{2\pi k_B T}{m}\right ).
\end{eqnarray}
The expression (\ref{e5}) can be obtained from Eq. (\ref{ffg}) by using Eq. (\ref{e4}). For an arbitrary barotropic equation of state $p(\rho)$ we have \cite{gen,nfp,longshort}:
\begin{eqnarray}
\label{e6}
F[\rho,{\bf u}]=\int \rho \frac{{\bf u}^2}{2}\, d{\bf r}+{1\over 2}\int\rho\Phi \, d{\bf r}+\int\rho\int^{\rho}\frac{p(\rho')}{{\rho'}^2}\, d{\bf r},
\end{eqnarray}
up to an additional constant.  A simple calculation gives \cite{nfp}:
\begin{equation}
\label{e7}
\dot F=-\int\xi\rho {\bf u}^2\, d{\bf r}\le 0.
\end{equation}
Therefore, $\dot F\le 0$ and $\dot F=0$ if, and only if, ${\bf u}={\bf 0}$ and $\nabla p+\rho\nabla\Phi={\bf 0}$ (hydrostatic equilibrium). In the isothermal case, this leads to the
mean field Boltzmann distribution (\ref{cp4}) with the temperature of the bath $T$.  Because of the $H$-theorem, the system converges, for $t\rightarrow +\infty$, towards a distribution that is a (local) minimum of free energy
at fixed mass. If several minima exist,
the selection depends on a notion of basin of attraction.  The
relaxation time is $t_{B}\sim 1/\xi$.

\subsection{The dispersion relation and the stability criterion}

The dispersion relation associated with the linearized mean field damped Euler
equations may be written as \cite{lr}:
\begin{equation}
\omega^2+i\xi\omega-\omega_0^2(k)=0,
\label{e8}
\end{equation}
with $\omega_0^2(k)=c_s^2k^2+(2\pi)^d\hat{u}(k)k^2\rho$. The complex pulsations are given by
\begin{equation}
\omega=\frac{-i\xi\pm\sqrt{4\omega_0^2(k)-\xi^2}}{2}.
\label{e9}
\end{equation}
The system is stable with respect to a perturbation with wavenumber $k$ when
\begin{equation}
c_s^2+(2\pi)^d\hat{u}(k)\rho>0,
\label{e10}
\end{equation}
and unstable otherwise \cite{lr}. This
stability criterion can also be obtained from the study of the second
order variations of the free energy (see Appendix \ref{sec_so}).

\subsection{Application to the BMF model,  self-gravitating systems, and plasmas}

\begin{figure*}
\centering
\includegraphics[width=0.5\textwidth]{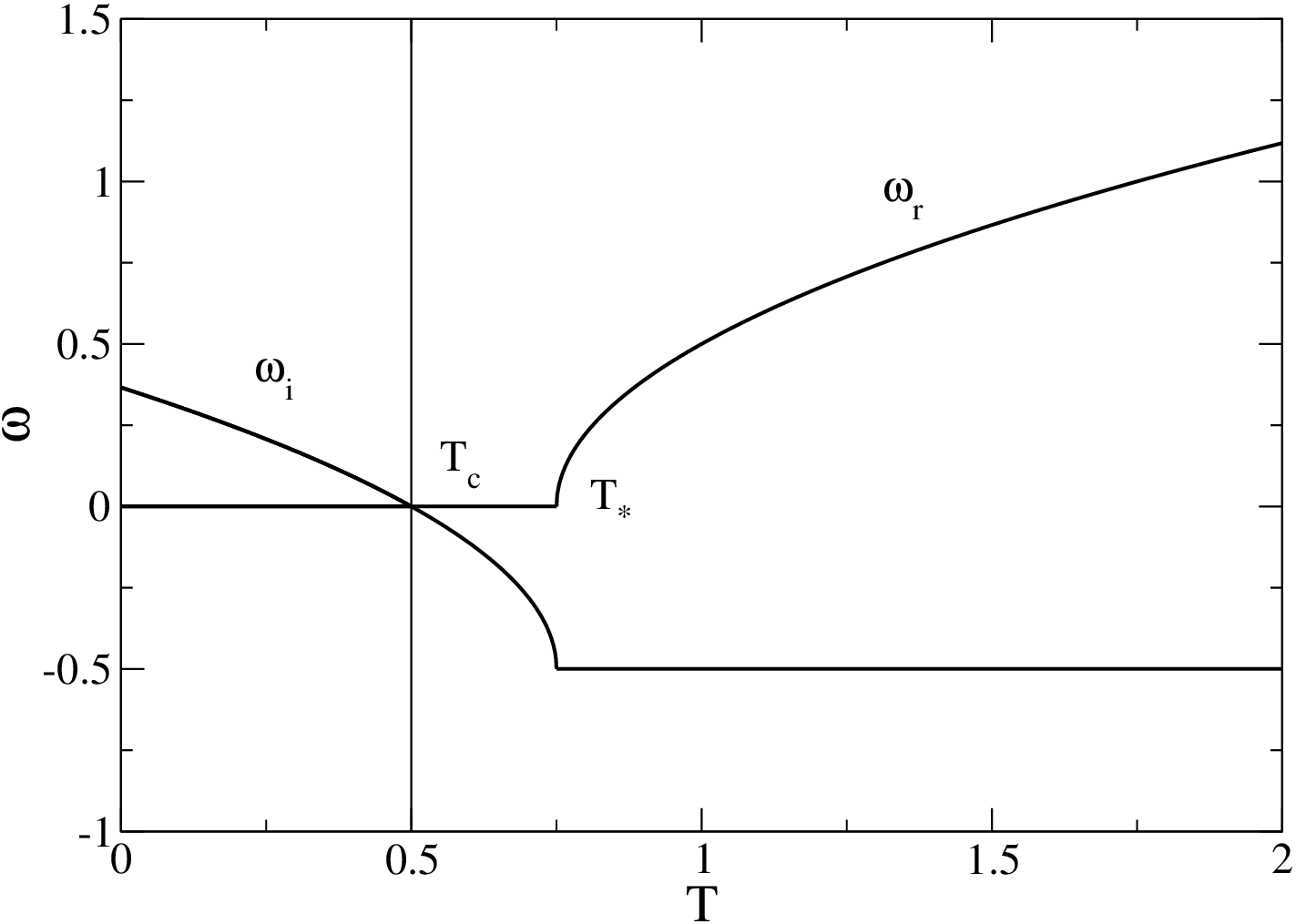}
\caption{Real and imaginary parts of the complex pulsation $\omega$ as a function of the temperature $T$ for the attractive BMF model described by the mean field damped Euler equations (we have taken $\xi=1$).}
\label{dampedEulerATTRACTIVE}
\end{figure*}

For the attractive BMF model, using $\hat{u}_n=\frac{1}{2N}(2\delta_{n,0}-\delta_{n,1}-\delta_{n,-1})$ and $\rho=1/(2\pi)$,  the dispersion relation (\ref{e8}) can be written as
\begin{equation}
\omega(\omega+i\xi)=c_s^2n^2,\qquad (n\neq \pm 1),\qquad \omega(\omega+i\xi)=c_s^2-\frac{1}{2}, \qquad (n= \pm 1).
\label{e13}
\end{equation}
For $n\neq \pm 1$, the complex pulsations are given by
\begin{equation}
\omega=\frac{-i\xi\pm\sqrt{4c_s^2n^2-\xi^2}}{2}.
\label{e15}
\end{equation}
If $n^2<\xi^2/4c_s^2$,  the perturbation decays exponentially rapidly at a rate $\omega_i=-\xi/2\pm \sqrt{\xi^2-4c_s^2n^2}<0$ without oscillating ($\omega_r=0$). If $n^2>\xi^2/4c_s^2$,  the perturbation oscillates with a pulsation $\omega_r=\pm (1/2)\sqrt{4c_s^2n^2-\xi^2}$ and is damped at a rate $\omega_i=-\xi/2<0$. For $n=\pm 1$, the complex pulsations are given by
\begin{equation}
\omega=\frac{-i\xi\pm\sqrt{4(c_s^2-\frac{1}{2})-\xi^2}}{2}.
\label{e16}
\end{equation}
The system is stable if $c_s^2>(c_s^2)_c=1/2$ and unstable  
if $c_s^2<1/2$. When $c_s^2<(c_s^2)_*=\xi^2/4+1/2$, we find  $\omega_r=0$ and
$\omega_i=-\xi/2\pm (1/2)\sqrt{\xi^2-4(c_s^2-1/2)}$. We have to distinguish two
cases. If $c_s^2<1/2$,  the perturbation grows exponentially rapidly at a rate
$\omega_i^{(+)}>0$ (unstable) without oscillating. If $c_s^2>1/2$,  the
perturbation decays exponentially rapidly at a rate $\omega_i^{(\pm)}<0$
(stable) without oscillating. When $c_s^2>\xi^2/4+1/2$ the perturbation
oscillates with a pulsation $\omega_r=\pm (1/2)\sqrt{4(c_s^2-1/2)-\xi^2}$ and is
damped at a rate $\omega_i=-\xi/2<0$. For the Euler equation ($\xi=0$), the
dispersion relation reduces to  $\omega^2=c_s^2n^2$ for $n\neq \pm 1$ and to
$\omega^2=c_s^2-1/2$ for $n=\pm 1$. The modes $n\neq \pm 1$ oscillate with a
pulsation $\omega_r=\pm c_s n$. For the modes $n=\pm 1$, we have to distinguish
two cases. If $c_s^2<1/2$, the perturbation grows exponentially rapidly at a
rate $\omega_i=\sqrt{1/2-c_s^2}>0$ without oscillating. If $c_s^2>1/2$, the
perturbation oscillates with a pulsation $\omega_r=\pm\sqrt{c_s^2-1/2}$. These
results are illustrated in Fig. \ref{dampedEulerATTRACTIVE}.

For self-gravitating systems, using $(2\pi)^d \hat{u}(k)=-S_d G/k^2$, the dispersion relation (\ref{e8}) can be written as
\begin{equation}
\omega(\omega+i\xi)=c_s^2k^2-\omega_G^2.
\label{e11}
\end{equation}
The system is stable if $k>k_J$ and unstable if $k<k_J$. The complex pulsations are given by
\begin{equation}
\frac{\omega}{\omega_G}=\frac{-\frac{i}{\cal N}\pm\sqrt{4(\frac{k^2}{k_J^2}-1)-\frac{1}{{\cal N}^2}}}{2}.
\label{e12}
\end{equation}
When $(k/k_J)^2<1/(4{\cal N}^2)+1$ we find $\omega_r=0$ and
$\omega_i/\omega_G=-1/(2{\cal N})\pm (1/2)\sqrt{1/{\cal N}^2-4(k^2/k_J^2-1)}$.
We have to distinguish two cases. If $k<k_J$,  the perturbation grows
exponentially rapidly at a rate $\omega_i^{(+)}>0$ (unstable) without
oscillating. If $k>k_J$,  the perturbation decays exponentially rapidly at a
rate $\omega_i^{(\pm)}<0$ (stable) without oscillating. When
$(k/k_J)^2>1/(4{\cal N}^2)+1$, the perturbation oscillates with a pulsation
$\omega_r/\omega_G=\pm (1/2)\sqrt{4(k^2/k_J^2-1)-1/{\cal N}^2}$ and is damped at
a rate $\omega_i/\omega_G=-1/(2{\cal N})<0$.
For the Euler equation ($\xi=0$), the dispersion relation reduces to  $\omega^2=c_s^2k^2-S_d G\rho$. If $k<k_J$, the perturbation grows exponentially rapidly at a rate $\omega_i/\omega_G=\sqrt{1-k^2/k_J^2}>0$ without oscillating. If $k>k_J$, the perturbation oscillates with a pulsation $\omega_r/\omega_G=\pm\sqrt{k^2/k_J^2-1}$.

\begin{figure*}
\centering
\includegraphics[width=0.5\textwidth]{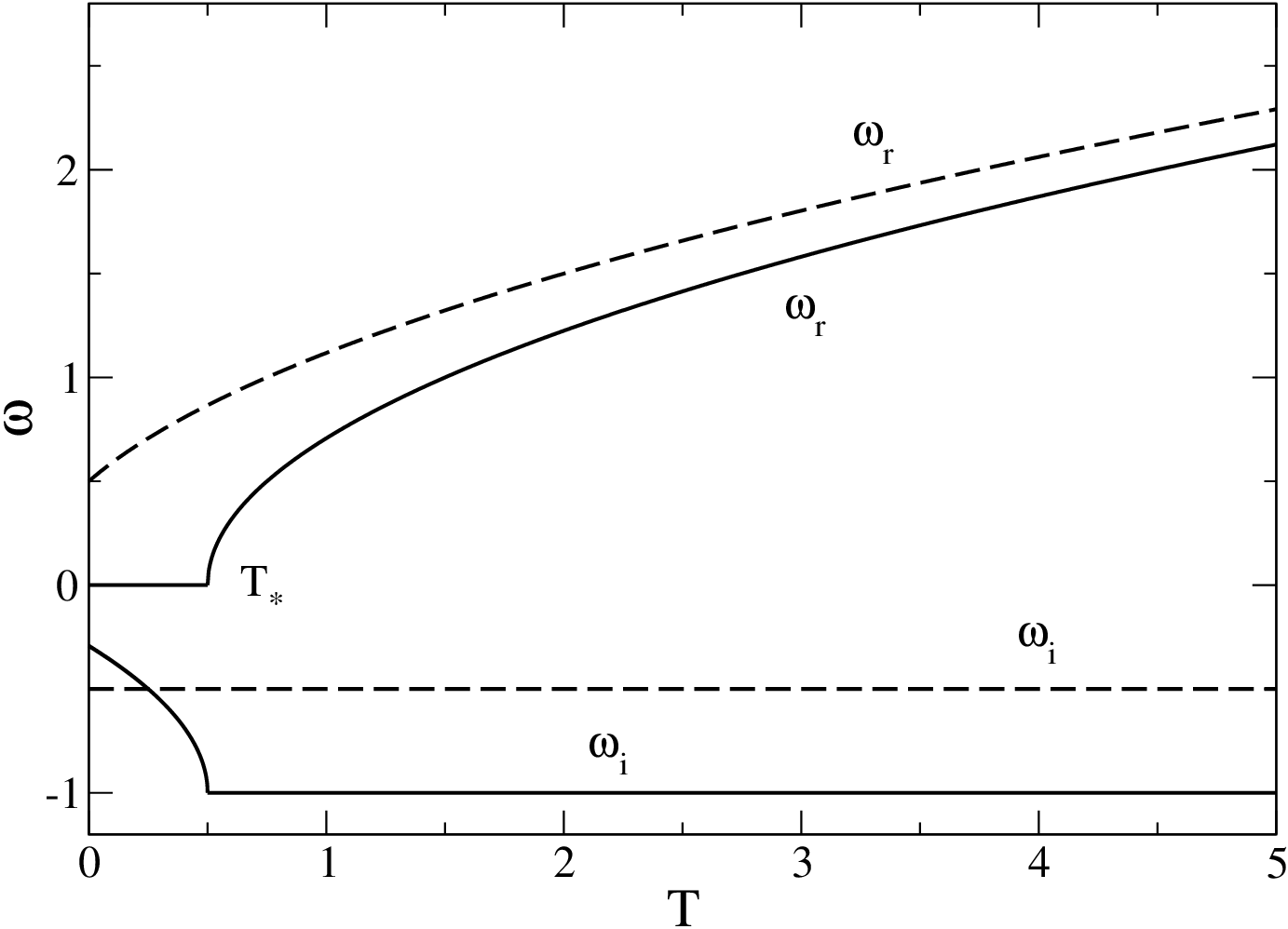}
\caption{Real and imaginary parts of the complex pulsation $\omega$ as a function of the temperature $T$ for the repulsive BMF model described by the mean field damped Euler equations. The dashed lines correspond to $\xi<\xi_c=\sqrt{2}$  (we have taken $\xi=1$) and the solid lines correspond to $\xi>\xi_c=\sqrt{2}$  (we have taken $\xi=2$).}
\label{dampedEulerREPULSIVE}
\end{figure*}

For the repulsive BMF model, using
$\hat{u}_n=-\frac{1}{2N}(2\delta_{n,0}-\delta_{n,1}-\delta_{n,-1})$
and $\rho=1/(2\pi)$, the dispersion relation (\ref{e8}) can be written as Eq.
(\ref{e13}-a) for $n\neq \pm 1$ and as
\begin{equation}
\omega(\omega+i\xi)=c_s^2+\frac{1}{2},
\label{e19}
\end{equation}
for $n= \pm 1$. The system is always stable. The discussion of the modes $n\neq \pm 1$ is the same as the one given previously so we consider here the modes $n=\pm 1$. The complex pulsations are given by
\begin{equation}
\omega=\frac{-i\xi\pm\sqrt{4(c_s^2+\frac{1}{2})-\xi^2}}{2}.
\label{e20}
\end{equation}
If $\xi<\xi_c=\sqrt{2}$, the perturbation oscillates with a pulsation
$\omega_r=\pm (1/2)\sqrt{4(c_s^2+1/2)-\xi^2}$ and is damped at a rate
$\omega_i=-\xi/2<0$.  We now assume $\xi>\sqrt{2}$. If
$c_s^2<(c_s^2)_*=\xi^2/4-1/2$ the perturbation is damped at a rate
$\omega_i=-\xi/2\pm (1/2)\sqrt{\xi^2-4(c_s^2+1/2)}<0$ without
oscillating ($\omega_r=0$). If $c_s^2>\xi^2/4-1/2$ the perturbation
oscillates with a pulsation $\omega_r=\pm
(1/2)\sqrt{4(c_s^2+1/2)-\xi^2}$ and is damped at a rate
$\omega_i=-\xi/2<0$. For the Euler equation ($\xi=0$), the dispersion
relation reduces to $\omega^2=c_s^2+1/2$. The perturbation oscillates
with a pulsation $\omega_r=\pm\sqrt{c_s^2+1/2}$. These results are
illustrated in
Fig. \ref{dampedEulerREPULSIVE}.

For Coulombian plasmas, using $(2\pi)^d \hat{u}(k)=S_d e^2/m^2k^2$, the dispersion relation (\ref{e8}) can be written as
\begin{equation}
\omega(\omega+i\xi)=c_s^2k^2+\omega_P^2.
\label{e17}
\end{equation}
The system is always stable. The complex pulsations are given by
\begin{equation}
\frac{\omega}{\omega_P}=\frac{-\frac{i}{\cal N}\pm\sqrt{4(\frac{k^2}{k_D^2}+1)-\frac{1}{{\cal N}^2}}}{2}.
\label{e18}
\end{equation}
If ${\cal N}>1/2$, the perturbation oscillates with a pulsation $\omega_r/\omega_P=\pm (1/2)\sqrt{4(k^2/k_D^2+1)-1/{\cal N}^2}$ and is damped at a rate $\omega_i/\omega_P=-1/(2{\cal N})<0$.
We now assume ${\cal N}<1/2$. If $(k/k_D)^2<1/(4{\cal N}^2)-1$ the perturbation is damped at a rate  $\omega_i/\omega_P=-1/(2{\cal N})\pm (1/2)\sqrt{1/{\cal N}^2-4(k^2/k_D^2+1)}<0$ without oscillating ($\omega_r=0$). If $(k/k_D)^2>1/(4{\cal N}^2)-1$ the perturbation oscillates with a pulsation $\omega_r/\omega_P=\pm (1/2)\sqrt{4(k^2/k_D^2+1)-1/{\cal N}^2}$ and is damped at a rate $\omega_i/\omega_P=-1/(2{\cal N})<0$. For the Euler equation ($\xi=0$), the dispersion relation reduces to  $\omega^2=c_s^2k^2+\omega_P^2$. The perturbation oscillates with a pulsation $\omega_r/\omega_P=\pm\sqrt{k^2/k_D^2+1}$.

Comparing the results of this section with the results of
Sec. \ref{sec_k}, we see that the complex pulsations associated with
the linearized mean field damped Euler equations are different from
the complex pulsations associated with the linearized mean field
Kramers equation except for $T=0$ (for the BMF  model) or for $k=0$ 
(for self-gravitating systems and plasmas). On the other hand, for the
attractive 
BMF model and for self-gravitating systems, the perturbation oscillates at high
$T$ or $k$ (for $\xi<+\infty$) contrary to the case of the mean field Kramers
equation where it is purely damped.

\section{Conclusion}

In this paper, we have solved the initial value problem for the
linearized mean field Kramers equation. The corresponding dielectric
function has been expressed in terms of incomplete Gamma
functions. Although the complex pulsations depend on the friction
coefficient, the stability criterion does not depend on it. As an
illustration, we have considered the attractive and repulsive BMF
models, self-gravitating systems, and plasmas. Previously known
results valid for the Vlasov equation (no friction $\xi=0$) and for
the mean field Smoluchowski equation (strong frictions $\xi\rightarrow
+\infty$) have been recovered as particular limits of the present
study. For $\xi>0$, the Landau damping is erased by frictional
effects.  We have also considered the damped mean field Euler
equations which include a dissipation term and for which the
dispersion relation can be solved analytically. Other applications and
extensions will be considered in future works.

\appendix

\section{The functions $F(\alpha,x)$ and $G(x)$}
\label{sec_f}

The function $F(\alpha,x)$ is defined in terms of incomplete Gamma
functions in Eq. (\ref{k13}). Using $\gamma(\alpha,x)\sim
e^{-x}x^{\alpha}/\alpha$ for $\alpha\rightarrow +\infty$, we find that
$F(\alpha,x)\sim x/\alpha^2$ for $\alpha\rightarrow +\infty$. Another
expression of this function in the form of a series is
\begin{equation}
F(\alpha,x)={e^x}\sum_{n=0}^{+\infty}\frac{(-1)^n}{n!}(x+n)\frac{x^n}{n+\alpha}.
\label{k14}
\end{equation}
We note that $F(\alpha,x)\rightarrow \pm\infty$ when $\alpha\rightarrow -n$ where $n\ge 0$ is any positive integer.  The function $F(\alpha,x)$ can also be written as 
\begin{equation}
F(\alpha,x)=\sum_{n=1}^{+\infty}\frac{n x^n}{\alpha(\alpha+1)...(\alpha+n)}=\Gamma(\alpha)\sum_{n=1}^{+\infty}\frac{nx^n}{\Gamma(\alpha+n+1)}.
\label{k14b}
\end{equation}
Using $\gamma(\alpha+1,x)=\alpha\gamma(\alpha,x)-x^{\alpha}e^{-x}$ we have
\begin{equation}
F(\alpha,x)=1+\frac{e^x}{x^{\alpha}}(x-\alpha)\gamma(\alpha,x).
\label{k14w}
\end{equation}
From Eq. (\ref{k14}) or from Eq. (\ref{k14w}), we directly obtain $F(x,x)=1$. On
the other hand, for $\epsilon\ll 1$ we can make the approximation 
$F(x+\epsilon,x)\simeq 1-\epsilon\, G(x)$ where we have defined
\begin{equation}
G(x)= e^x\sum_{n=0}^{+\infty}\frac{(-1)^n}{n!}\frac{x^n}{n+x}.
\label{f1}
\end{equation}
This function may also be written as
\begin{equation}
G(x)= \frac{e^x}{x^{1+x}}\left\lbrack \Gamma(1+x)-x\Gamma(x,x)\right\rbrack,
\label{f2}
\end{equation}
where
\begin{equation}
\Gamma(\alpha,x)=\int_x^{+\infty}t^{\alpha-1}e^{-t}\, dt, \qquad \Gamma(\alpha)=\int_0^{+\infty}t^{\alpha-1}e^{-t}\, dt,
\label{f3}
\end{equation}
are the incomplete and complete Gamma functions. We have the asymptotic behaviors
\begin{equation}
G(x)\sim \frac{1}{x} \qquad (x\rightarrow 0),\qquad G(x)\sim\sqrt{\frac{\pi}{2x}} \qquad (x\rightarrow +\infty).
\label{f4}
\end{equation}
In order to obtain the second behavior, we have used the results
\begin{equation}
\Gamma(x,x)\sim x^{x-1}e^{-x}\sqrt{\frac{\pi x}{2}}, \qquad \Gamma(1+x)\sim \sqrt{2\pi x}\left (\frac{x}{e}\right )^x, \qquad (x\rightarrow +\infty).
\label{f5}
\end{equation}

\section{Thermodynamical stability of the mean field Maxwell-Boltzmann distribution}
\label{sec_so}

The steady states of the mean field Kramers equation
(\ref{browbbgky7}) correspond to the mean field Maxwell-Boltzmann distribution
(\ref{mba1}). They are the critical points of the free energy
(\ref{ffg}) at fixed mass. Using general arguments based on the
fact that the free energy is the Lyapunov functional of the mean
field Kramers equation, we can
show that dynamical and thermodynamical stability coincide \cite{nfp}:
the mean field Maxwell-Boltzmann distribution is dynamically stable with
respect to the Kramers equation if, and only if, it is a (local)
minimum of free energy at fixed mass (thermodynamical stability).

To solve the minimization problem
\begin{eqnarray}
\label{nff1}
F(T)=\min_{f}\left\lbrace F\lbrack f\rbrack=E[f]-T S[f]\, |\, M\lbrack f\rbrack=M\right\rbrace,
\end{eqnarray}
we can proceed in two steps (see, e.g., Appendix A of \cite{cdpre}). We first minimize $F[f]$ at fixed normalization {\it and} density $\rho({\bf r})$. This gives
\begin{eqnarray}
\label{aace2o}
f({\bf r},{\bf v})=\left (\frac{\beta m}{2\pi}\right )^{d/2}\rho({\bf r}) e^{-\beta m v^2/2}.
\end{eqnarray}
Using Eq. (\ref{aace2o}) we can express the free energy $F[f]$ given
by Eq.  (\ref{ffg}) as a functional of the density $\rho$. This leads to
Eq. (\ref{tsce1zero}). Finally, the solution of the minimization problem
(\ref{nff1}) is given by Eq. (\ref{aace2o}) where $\rho({\bf r})$ is
the solution of the minimization problem
\begin{eqnarray}
\label{ff1}
F(T)=\min_{\rho}\left\lbrace F\lbrack \rho\rbrack\, |\, M\lbrack \rho\rbrack=M\right\rbrace.
\end{eqnarray}
It can be shown that the minimization problems (\ref{nff1}) and
(\ref{ff1}) are equivalent for global and local minimization
\cite{cdpre}. If we consider the overdamped model, we immediately arrive at the minimization problem (\ref{ff1}).

The critical points of (\ref{ff1}) at fixed mass satisfy $\delta F+\alpha T\delta M=0$ and they lead to the mean field Boltzmann distribution (\ref{cp4}). The second order variations of free energy are given by
\begin{eqnarray}
\label{tsce4}
\delta^{2} F={1\over 2}\int \delta\rho\delta\Phi \, d{\bf r}+\frac{k_B T}{m}\int {(\delta\rho)^{2}\over 2\rho}\, d{\bf r},
\end{eqnarray}
with
\begin{equation}
\label{ejl3}
\delta\Phi({\bf r})=\int u({\bf r}-{\bf r}') \delta\rho({\bf r}')\, d{\bf r}'.
\end{equation}
The Boltzmann distribution is a (local) minimum of free energy at fixed mass if,
and only if, $\delta^2 F>0$ for all perturbations satisfying $\int \delta\rho\,
d{\bf r}=0$.  If the critical point of free energy is 
spatially homogeneous, we can decompose the perturbation $\delta\rho$ in Fourier
modes as in Eq. (\ref{g10}). The second variations of free energy can
then  be rewritten as
\begin{eqnarray}
\delta^2 F=\frac{(2\pi)^d}{2\rho}\int \left\lbrack \frac{k_B T}{m}+(2\pi)^d \hat{u}(k)\rho\right\rbrack |\delta\hat{\rho}({\bf k})|^2\, d{\bf k}.
\label{dyn1a}
\end{eqnarray}
If $\hat{u}(k)>0$ for all $k$ (repulsive interaction),  
the homogeneous phase is thermodynamically stable. If  $\hat{u}(k)<0$ for some
mode(s) $k$ (attractive interaction), the homogeneous phase is thermodynamically
 stable when $T>T_c=(\rho m/k_B)(2\pi)^d\max_k|\hat{u}(k)|$ and
thermodynamically unstable (with respect to the modes such that $k_BT/m+(2\pi)^d
\hat{u}(k)\rho<0$) when $T<T_c$. This returns the stability criterion
(\ref{k19}).

For the Vlasov equation, corresponding to Eq. (\ref{browbbgky7}) with $\xi=0$,
the 
functional (\ref{ffg}) is conserved. It may be interpreted as an energy-Casimir
functional. A minimum of this functional is formally nonlinearly dynamically
stable with respect to the Vlasov equation \cite{holm}. In general, this
criterion provides just a sufficient condition of dynamical stability. More
refined dynamical stability criteria exist (see \cite{ccstab} for details).
However, for spatially homogeneous distributions, this criterion can be shown to
be both necessary and sufficient \cite{ccstab}. This leads to Eq. (\ref{dyn1a})
then to Eq. (\ref{v14}). For distribution functions different from the Maxwell
distribution, these results remain valid for the Vlasov equation provided that
$k_B T/m$ is replaced by $c_s^2$ where $c_s$ is the velocity of sound in the
``corresponding barotropic gas'' (see \cite{cvb,cd,nyquistgrav} for details).

For the mean field damped Euler equations (\ref{e1})-(\ref{e3}), the free energy
(\ref{e5}) plays the role of a Lyapunov functional. Therefore, a steady
state of the mean field damped Euler equations  is dynamically stable if, and
only, if it is a (local) minimum of free energy at fixed mass (thermodynamical
stability). This leads to the minimization problem (\ref{ff1}) then to Eq.
(\ref{dyn1a}) where $k_B T/m$ is replaced by $c_s^2$, and finally to the
stability criterion (\ref{e10}). For the mean field Euler equation
($\xi=0$), the functional (\ref{e5}) is conserved. It corresponds to the
energy functional of a barotropic gas \cite{cvb,cd,nyquistgrav}. A minimum of
this functional at fixed mass is formally nonlinearly dynamically stable with
respect to the mean field Euler equations \cite{holm}. This leads  to the
stability criterion (\ref{e10}). Since the energy
functional and the mass are the only conserved quantities, this criterion
provides a necessary and sufficient condition of dynamical stability.

\section{An alternative calculation of $w(x)$}
\label{sec_alt}

The Fourier-Laplace transform of the polarization function associated with the linearized Vlasov equation is given by Eq. (\ref{v4}-a). Taking its inverse Laplace transform and using the Cauchy residue theorem, we get the expression (\ref{v4b}). Taking the Laplace transform of this expression and using Eq. (\ref{g24}), we find that the dielectric function can be written as
\begin{equation}
\label{alt1}
\epsilon(k,\omega)=1+(2\pi)^d\hat{u}(k)\rho k^2\int_0^{+\infty} dt\, e^{i\omega t} t e^{-\frac{k^2 t^2}{2\beta m}}.
\end{equation}
With the change of variables $x=k t/\sqrt{\beta m}$, the foregoing equation may be rewritten as Eq. (\ref{v9}) with 
\begin{equation}
\label{alt2}
W(z)=\int_0^{+\infty} dx\, e^{izx}xe^{-x^2/2}.
\end{equation}
Assuming that $z=ix$ where $x$ is real, integrating by parts, and using the identity
\begin{equation}
\label{alt3}
\int_0^{+\infty} dx\, e^{-\gamma x} e^{-\frac{x^2}{4\beta}}=\sqrt{\pi\beta}e^{\beta \gamma^2}{\rm erfc}(\gamma\sqrt{\beta}),
\end{equation}
we obtain $W(ix)=w(x/\sqrt{2})$ where $w(x)$ is given by Eq. (\ref{v16}).

\section{The expansion $\xi\rightarrow 0$}
\label{sec_xizero}

According to Eqs. (\ref{g24}) and (\ref{k9}), the dielectric function associated with the linearized mean field Kramers equation may be written as
\begin{equation}
\label{xizero1}
\epsilon({k},\omega)=1+\frac{1}{\xi}(2\pi)^d\hat{u}(k)\rho k^2\int_{0}^{+\infty} e^{i\omega t}(1-e^{-\xi t})e^{-\frac{Dk^2}{\xi^2}t}e^{\frac{Dk^2}{\xi^3}(1-e^{-\xi t})}\, dt.
\end{equation}
Expanding the integrand for $\xi\rightarrow 0$, we find that the dispersion relation at the order $O(\xi)$ is   
\begin{equation}
\label{xizero2}
1+(2\pi)^d\hat{u}(k)\rho\beta m \left\lbrack W\left (\sqrt{\beta m}\frac{\omega}{k}\right )+\xi\frac{\sqrt{\beta m}}{k}U\left (\sqrt{\beta m}\frac{\omega}{k}\right )\right\rbrack=0,
\end{equation}
where $W(z)$ is given by Eq. (\ref{alt2}) and $U(z)$ by
\begin{equation}
\label{xizero3}
U(z)=\int_0^{+\infty} dx\, e^{izx}x\left (-\frac{1}{2}x
+\frac{1}{6}x^3\right ) e^{-x^2/2}.
\end{equation}
These functions may be rewritten as
\begin{equation}
\label{xizero4}
W(z)=1-\sqrt{\frac{\pi}{2}}z e^{-\frac{z^2}{2}}\left \lbrack -i+{\rm erfi}\left (\frac{z}{\sqrt{2}}\right )\right \rbrack,
\end{equation}
\begin{equation}
\label{xizero5}
U(z)=\frac{1}{12}z\left \lbrace 2i(2-z^2)+\sqrt{2\pi}z(z^2-3)e^{-\frac{z^2}{2}}\left\lbrack  1+i\, {\rm erfi}\left (\frac{z}{\sqrt{2}}\right )\right \rbrack\right\rbrace,
\end{equation}
where ${\rm erfi}(z)={\rm erf}(iz)/i$. We note that
$U(z)=-\frac{i}{6}z\lbrack1+(z^2-3)W(z)\rbrack$. 

Considering the attractive BMF model, assuming
$\omega=i\omega_i$ and taking the limit $T\rightarrow 0$, we
obtain after careful calculations to order $O(T)$:
\begin{equation}
\label{xizero7}
\omega_i=\frac{1}{\sqrt{2}}-\frac{\xi}{2}+\left (-\frac{3}{\sqrt{2}}+4\xi\right )T\qquad (T\rightarrow 0).
\end{equation}
We can check that this result agrees with Eq. (\ref{k23a}) at the order $O(\xi)$. On the other hand, taking the limit $T\rightarrow T_c$ and using $W(z)\simeq 1+i\sqrt{\pi/2}\, z+...$ and $U(z)\sim
iz/3$ for $z\rightarrow 0$, we get
\begin{equation}
\label{xizero6}
\omega_i=\frac{1}{\sqrt{\pi}}\left (1-\frac{2\xi}{3\sqrt{\pi}}+...\right )(1-2T), \qquad (T\rightarrow T_c).
\end{equation}
Using $G(x)\simeq \sqrt{\pi/(2x)}+1/(3x)+...$ for $x\rightarrow
+\infty$, we can check that Eq. (\ref{xizero6}) agrees with
Eq. (\ref{k23b}) at the order $O(\xi)$. 

Considering the repulsive BMF model, and taking the limit $T\rightarrow 0$ for which $\omega_i\ll\omega_r$, we obtain after careful
calculations to order $O(T)$:
\begin{equation}
\label{xizero8}
\omega_r^2=\frac{1}{{2}}+3T-\frac{\xi}{T^{7/2}}\frac{1}{96}\sqrt{\frac{\pi}{2}}e^{-\frac{1}{4T}}\qquad (T\rightarrow 0),  
\end{equation}
\begin{equation}
\label{xizero9}
\omega_i=-\frac{1}{8}\sqrt{\frac{\pi}{2}}\frac{1}{T^{3/2}}e^{-\frac{1}{4T}}-\frac{\xi}{2}(1+8T)-\frac{\xi}{T^{6}}\frac{\pi}{3072}e^{-\frac{1}{2T}}\qquad (T\rightarrow 0). 
\end{equation}
We can check that this result agrees with Eq. (\ref{grle}) at the
order $O(\xi)$ except for the exponentially small terms
(see footnote
4). When $\xi=0$ we recover the Landau damping (\ref{v38}) but as soon
as $\xi>0$, the Landau damping for $T\rightarrow 0$ becomes
negligible (subdominant) with respect to the frictional terms.

The dispersion relation (\ref{xizero2}) can also be solved
perturbatively by writing $\omega=\omega_0+\xi\omega_1+...$ with $\xi\ll
1$. Substituting this expansion in Eq. (\ref{xizero2}) we find that $\omega_0$ is given by Eq. (\ref{v12}) and that
\begin{equation}
\label{fin1}
\omega_1=-\frac{U\left (\sqrt{\beta m}\frac{\omega_0}{k}\right )}{W'\left (\sqrt{\beta m}\frac{\omega_0}{k}\right )}.
\end{equation}
Using the identity $W'(z)=(1/z-z)W(z)-1/z$ and expressing $U(z)$ in terms of $W(z)$ we obtain
\begin{equation}
\label{fin2}
1+(2\pi)^d\hat{u}(k)\rho\beta m W(z_0)=0, \qquad \omega_1=i\frac{z_0^2}{6}\frac{1+(z_0^2-3)W(z_0)}{(1-z_0^2)W(z_0)-1},
\end{equation}
where we have defined $z_0=\sqrt{\beta m}\omega_0/k$. These results can be obtained by other methods (in preparation). For the attractive and repulsive BMF models, we get
\begin{equation}
\label{fin3}
1\mp \frac{1}{2T}W\left (\frac{\omega_0}{\sqrt{T}}\right )=0, \qquad \omega_1=i\frac{\omega_0^2}{6T}\frac{1\pm 2(\omega_0^2-3T)}{\pm 2(T-\omega_0^2)-1}.
\end{equation}
For $T\rightarrow 0$ and $T\rightarrow T_c$ (in the attractive case), we recover Eqs. (\ref{xizero7})-(\ref{xizero9}). For $T\rightarrow +\infty$ we find
\begin{equation}
\label{fin4}
\omega_i\simeq -\sqrt{2T\ln T}+\frac{1}{3}\xi\ln T,
\end{equation} 
in the attractive case and
\begin{equation}
\label{fin5}
\omega_r\simeq \pi\sqrt{\frac{T}{2\ln T}}-\frac{\pi}{3}\xi,\qquad \omega_i=-\sqrt{2T\ln T}+\frac{1}{3}\xi\ln T,
\end{equation}
in the repulsive case.

We have given here the results for the BMF model. The corresponding
results for self-gravitating systems and plasmas can be obtained by
using the correspondences of Sec. \ref{sec_cor}.


\begin{thebibliography}{}

\bibitem{houches} {\small {\it Dynamics and thermodynamics of systems
with long range interactions}, edited by T. Dauxois {\it et al.},
Lecture Notes in Physics {\bf 602}, (Springer, 2002)}
\bibitem{assisebook} {\small {\it Dynamics and thermodynamics of systems
with long range interactions: Theory and experiments}, edited by
A. Campa {\it et al.}, AIP Conf. Proc. {\bf 970} (AIP, 2008). }
\bibitem{oxford}  {\small  {\it Long-Range Interacting Systems}, edited by T. Dauxois, S. Ruffo and L. Cugliandolo, Les Houches Summer School 2008, (Oxford: Oxford University Press, 2009)}
\bibitem{cdr}  {\small A. Campa, T. Dauxois, S. Ruffo,   Physics Reports {\bf 480}, 57 (2009)}
\bibitem{balescubook}{\small  R. Balescu, {\it Statistical Mechanics of Charged Particles} (Wiley, 1963)}
\bibitem{bt}  {\small  J. Binney, S. Tremaine,
{\it Galactic Dynamics} (Princeton Series in Astrophysics, 1987)}
\bibitem{paddy}  {\small  T. Padmanabhan, Phys. Rep. {\bf 188}, 285 (1990)}
\bibitem{newton}  {\small P.K. Newton, {\it The $N$-Vortex Problem: Analytical Techniques}, in {\it Applied Mathematical Sciences} {\bf 145} (Springer-Verlag, Berlin, 2001)}
\bibitem{chavhouches}  {\small P.H. Chavanis, {\it Statistical mechanics of two-dimensional vortices and stellar systems} in \cite{houches}.}
\bibitem{ijmpb}  {\small  P.H. Chavanis, Int. J. Mod. Phys. B {\bf 20}, 3113
(2006) }
\bibitem{bv}  {\small F. Bouchet, A. Venaille, Phys. Rep. {\bf 515}, 227 (2012)
}
\bibitem{ar}  {\small  M. Antoni, S.  Ruffo, Phys. Rev. E {\bf  52}, 2361 (1995)  }
\bibitem{landau}  {\small L. Landau,  Journ. of Phys. {\bf 10}, 25 (1946)}
\bibitem{vlasov1}  {\small A. Vlasov, Journ. Exper. a. Theor. Phys.  {\bf 8}, 291 (1938)}
\bibitem{vlasov2}  {\small A. Vlasov,  Journ. of Phys.  {\bf 9}, 25 (1945)}
\bibitem{villani}  {\small C. Mouhot, C. Villani, Acta Mathematica {\bf 207}, 29 (2011)}
\bibitem{lb62}  {\small D. Lynden-Bell, MNRAS {\bf 124}, 279 (1962)}
\bibitem{nyquistgrav}  {\small P.H. Chavanis,  Eur.
Phys. J. B {\bf 85}, 229 (2012)}
\bibitem{hb1}  {\small P.H. Chavanis, Physica A  {\bf 361}, 55 (2006)}
\bibitem{hb2}  {\small P.H. Chavanis, Physica A  {\bf 361}, 81 (2006)}
\bibitem{hb5}  {\small P.H. Chavanis, Physica A  {\bf 387}, 5716 (2008)}
\bibitem{longshort}  {\small P.H. Chavanis, Physica A {\bf 390}, 1546 (2011)}
\bibitem{virial1}  {\small P.H. Chavanis, C. Sire, Phys. Rev. E {\bf 73},
066103 (2006)}
\bibitem{ks}  {\small E. Keller, L.A. Segel, J. Theor. Biol. {\bf 26}, 399 (1970)}
\bibitem{sopikbec}  {\small J. Sopik, C. Sire, P.H. Chavanis, Phys. Rev. E {\bf
74}, 011112 (2006)}
\bibitem{cvb}  {\small P.H. Chavanis, J. Vatteville, F. Bouchet, Eur. Phys. J. B {\bf 46}, 61 (2005)}
\bibitem{cbo}  {\small P.H. Chavanis, F. Baldovin, E. Orlandini, Phys. Rev. E {\bf 83}, 040101(R) (2011)}
\bibitem{bco}  {\small F. Baldovin, P.H. Chavanis, E. Orlandini, Phys. Rev. E {\bf 79}, 011102 (2009)}
\bibitem{bmf}  {\small P.H. Chavanis, {\it The Brownian Mean Field model}
[arXiv:1306]}
\bibitem{risken}  {\small H. Risken, {\it The Fokker-Planck equation} (Springer, 1989)}
\bibitem{nfp}  {\small P.H. Chavanis, Eur. Phys. J. B {\bf 62}, 179 (2008)}
\bibitem{gen}  {\small  P.H. Chavanis, Phys. Rev. E {\bf 68}, 036108
(2003)}
\bibitem{spa}  {\small P.H. Chavanis, Physica A {\bf
340}, 57 (2004)}
\bibitem{cll}  {\small P.H. Chavanis, P. Lauren\c cot, M. Lemou, Physica A {\bf
341}, 145 (2004)}
\bibitem{gross}  {\small E.P. Gross, Phys. Rev. {\bf 158}, 146
(1967)}
\bibitem{lr}  {\small P.H. Chavanis, Eur. Phys. J. Plus {\bf 128}, 38 (2013)}
\bibitem{cd}  {\small P.H. Chavanis, L. Delfini, Eur. Phys. J. B {\bf 69}, 389 (2009)}
\bibitem{fried}  {\small B.D. Fried, S.D. Conte, {\it The Plasma Dispersion
Function} (Academic Press, New York, 1961)}
\bibitem{chandra}  {\small S. Chandrasekhar, Rev. Mod. Phys. {\bf 15}, 1 (1943)}
\bibitem{hydrobrown}  {\small P.H. Chavanis, Physica A  {\bf 389}, 375 (2010)}
\bibitem{cdpre}  {\small P.H. Chavanis, L. Delfini, Phys. Rev. E {\bf 81}, 051103 (2010)}
\bibitem{holm}  {\small D. Holm, J. Marsden, T. Ratiu, A. Weinstein, Phys. Rep.
{\bf 123}, 1 (1985)}
\bibitem{ccstab}  {\small A. Campa, P.H. Chavanis, J. Stat. Mech. {\bf 6},
06001 (2010)}



\end{thebibliography}
\end{document}